# Forecasting Strong Subsequent Earthquakes in Japan using an improved version of NESTORE Machine Learning Algorithm

Gentili S.[1*], Chiappetta G. D.[1,2], Petrillo G.[3,4,5], Brondi P.[1], Zhuang J.[3]

1) National Institute of Oceanography and Applied Geophysics - OGS, Via Treviso 55, 33100 Udine, Italy, sgentili@ogs.it, pbrondi@ogs.it
2) Università della Calabria, Via Pietro Bucci, 87036 Rende, Cosenza, Italy giuseppe.chiappetta@unical.it
3) The Institute of Statistical Mathematics – ISM, 10-3 Midori-cho, Tachikawa Tokyo 190-8562, Japan giuseppe51289@gmail.com, zhuangjc@ism.ac.jp
4) Scuola Superiore Meridionale – SSM, Via Marchese Campodisola, 16, 80133 Naples, Italy giuseppe.petrillo@ssmeridionale.it 5)
5) Earth Observatory of Singapore, Nanyang Technological University (NTU), 50 Nanyang Avenue, Singapore 639798 giuseppe.petrillo@ntu.edu.sg

corresponding author: Stefania Gentili sgentili@ogs.it

**Abstract**

The advanced machine learning algorithm NESTORE (Next STrOng Related Earthquake) was developed to forecast strong aftershocks in earthquake sequences and has been successfully tested in Italy, western Slovenia, Greece, and California. NESTORE calculates the probability of aftershocks reaching or exceeding the magnitude of the main earthquake minus one and classifies clusters as type A or B based on a 0.5 probability threshold.

In this study, NESTORE was applied to Japan using data from the Japan Meteorological Agency catalog (1973-2024). Due to Japan's high seismic activity and class imbalance, new algorithms were developed to complement NESTORE. The first is a hybrid cluster identification method using ETAS-based stochastic declustering and deterministic graph-based selection. The second, REPENESE (RElevant features, class imbalance PErcentage, NEighbour detection, SElection), is optimized for detecting outliers in skewed class distributions.

A new seismicity feature was proposed, showing good results in forecasting cluster classes in Japan. Trained with data from 1973 to 2004 and tested from 2005 to 2023, the method correctly forecasted 75% of A clusters and 96% of B clusters, achieving a precision of 0.75 and an accuracy of 0.94 six hours after the mainshock. It accurately classified the 2011 Tōhoku event cluster. Near-real-time forecasting was applied to the sequence after the April 17, 2024 M6.6 earthquake in Shikoku, classifying it as a "Type B cluster," with validation expected on October 31, 2024.

# 1. Introduction

Earthquakes often occur in clusters. When a strong earthquake occurs, the question arises as to the number, timing and strength of the subsequent aftershocks. Recent studies have shown that machine learning models are proving to be effective in forecasting the number and/or location of aftershocks following a major earthquake (DeVries et al., 2018; Lippiello et al., 2019a; Dascher-Cousineau et al., 2023; Schimmenti et al., 2024). One of the fundamental questions remains the forecast of the maximum aftershock magnitude, which still lacks a well-defined answer. Research focuses in particular on the value of Dm, where Dm equals Mm (the magnitude of the main earthquake) minus Ma (the magnitude of the strongest aftershock) (Saichev & Sornette, 2005; Zhuang & Ogata, 2006; Vere-Jones & Zhuang, 2008; Luo & Zhuang (2016); Shcherbakov et al. 2019; Gulia et al., 2020; Spassiani et al., 2024).

NESTORE (Next STrOng Related Earthquake - Gentili et al., 2023) is a machine learning algorithm that categorizes clusters of seismic activity into two types based on Dm: Type A when Dm is less than 1, and Type B when it is not, according to a classification first introduced by Vorobieva and Panza (1993). Type A clusters are particularly dangerous in densely populated areas, as seismic events of similar magnitude can repeatedly affect buildings and human activities, often with severe consequences. Therefore, the main objective of NESTORE is to estimate the probability that the following cluster will be of type A.

NESTORE uses an operational approach and is designed to be applied after a first strong earthquake, when it is not yet known whether this event is the mainshock or whether a stronger earthquake will follow. It replaces the concept of "mainshock" with the concept of "o-mainshock" (where "o" stands for operative, Gentili and Di Giovambattista, 2022), i.e. the first earthquake with a magnitude above a region and catalogue dependent threshold. Starting from the o-mainshock, the algorithm identifies the corresponding cluster and evaluates several seismic features to classify the cluster starting from a few hours after the o-mainshock. It is important to note that the o-mainshock is not necessarily the strongest earthquake in the cluster and can be followed by a stronger event, resulting in a negative Dm.

In this study, we apply the NESTORE machine learning algorithm to the Japan Meteorological Agency (JMA) catalog from 1973 (JMA 2024). Japan has a high seismicity rate and frequently experiences strong earthquakes, which often occur in clusters both spatially and temporally. This provides an extensive database to study the characteristics of seismicity and test the capabilities of the software.

The strongest in a series of significant earthquakes that have struck Japan over time, both inland and offshore, is the 2011 Tōhoku sequence, which began with a magnitude 7.3 event on March 9, 2011 and was followed two days later by a magnitude 9.1 event (an example of negative Dm). The numerous earthquakes in Japan are due to the complex geodynamic context in which the oceanic plates of the Pacific and Philippine Seas are subducting beneath the Eurasian plate (Wang and Zao, 2021 and references therein). This causes crustal, intermediate and very deep earthquakes, making Japan one of the most seismologically complex regions.

The application of a semi-automatic forecasting algorithm such as NESTORE in Japan is challenging; three main problems were encountered: (i) cluster identification, (ii) imbalance between Type A and Type B classes, and (iii) the presence of outliers.
Regarding the identification of clusters, the usual window-based methods fail in Japan due to the high seismicity rate and the proximity of many seismogenic structures. Clusters from different but close sources cannot be easily separated from each other, if they are close in time. In addition, the stress redistribution of a seismogenic zone can influence the neighboring zones, which leads to outliers in the machine learning training process.

Class imbalance poses another challenge, as a large prevalence of type B clusters complicates class discrimination compared to previous NESTORE applications in Italy, western Slovenia, Greece and California (Gentili and Di Giovambattista 2017, 2020, 2022; Anyfadi et al., 2023; Brondi et al., 2023, 2024). Indeed, this imbalance also complicates the automatic detection of outliers, especially for type A clusters, for which only a few instances are available.

In our study, we used NESTOREv1.0, a free software implementing the NESTORE algorithm (Gentili et al., 2023). Taking advantage of its modular structure, we substituted the cluster identification module with another algorithm we developed for this work based on a stochastic declustering approach derived by the ETAS model (Zhuang et al., 2002). In addition, we developed a new outlier detection method that takes into account class imbalancing. Finally, we proposed a new feature of seismicity for A and B type classes discrimination particularly suited for Japan.

The paper is structured as follows: Section 2 describes the seismotectonic environment and the dataset we used, Section 3 describes the new cluster detection algorithm, summarizes the NESTOREv1.0 algorithm and the newly proposed seismicity feature, and ends with the new outlier removal algorithm. Section 4 describes the results: we trained the software with the Japanese JMA catalog from 1973 to 2004 and tested it with data from 2005 to 2023; the first two subsections describe training and testing, the third describes the duration, productivity, and seismic moment for A and B clusters in Japan, while Section 4.4 shows an example of applying NESTORE in near real-time (6 hours after the mainshock) to a 2024 sequence. Section 5 contains a discussion and conclusions with lessons learned from the seismicity in Japan.

## 2. Japan seismotectonic and available data

### 2.1 Japanese Seismicity

The seismicity in Japan is predominantly driven by the convergence of several tectonic plates (see Fig 1). The Pacific plate subducts beneath the Eurasian plate along the Japan Trench, causing megathrust earthquakes. Notably, the Philippine Sea plate also subducts beneath the Eurasian plate in the southwestern region, generating significant interplate seismicity in the Nankai Trough. The subduction rate is high for both but with some differences: up to ~9-11 cm/year for the Pacific plate (Argus et al., 2011; DeMets et al., 2010; Romano et al., 2014) and ~6-11 cm/year for the Philippine Sea plate (e.g. Wu et al., 2016). Furthermore, the North American plate interacts with the Eurasian plate in the northern part of Japan, contributing to complex deformation and seismic activity. Also, Japan is renowned for its volcanic activity, with numerous active volcanoes scattered throughout the archipelago. Volcanic tremors and eruptions can be a source of seismic activity.

In this paper, we used the Japan Meteorological Agency (JMA) catalog. The database's accuracy and reliability have proven crucial for understanding seismic patterns, monitoring fault activities, and issuing real-time earthquake information. In the past, the JMA used mechanical-type strong motion sensors but later switched to digital recording-type electromotive sensors that could handle strong ground shaking. The JMA set up these sensors at various locations across Japan to enhance intensity observation. Currently, they use a network of about 670 measuring intensity meters and multifunctional seismic observation equipment for digital strong motion waveform observation. The JMA seismic catalog underwent significant improvements in 1997. Before this improvement, data collection and updates were handled independently by various organizations, as detailed at https://www.jishin.go.jp/resource/column/16spr_p6/. Now, the catalog can be easily downloaded from https://www.jma.go.jp/jma/.

The increased number of stations and the more homogeneous organization of the catalog also resulted in a decrease of the completeness magnitude until March 2011, when one of the strongest earthquakes of the last years struck Tōhoku. This sequence started with a Mw=7.3 event on March 9, 2011 and two days later a Mw=9.1 occurred, causing also a huge tsunami, which had a large toll in terms of human lives, environmental impact and economic costs, which in turn triggered the Fukushima Dai-ichi nuclear power plant accident, a large amount of released radionuclides (Kaizer et al., 2023) which further aggravated the situation with loss of radiations and other bad side effects for the population. The Tōhoku great event is a deadly example of the amount of energy that can be radiated when a plate boundary ruptures for ~400 km with an estimated slip of ~50m (e.g. Lay, 2018), but fortunately earthquakes like this are quite rare. However, more than 80 events with M≥7 occurred during the last 50 years (1973-2023) in the region comprising the sea of Japan, the Japan inland, Taiwan and along the two main subduction trenches. The aftershocks of the sequence dominated the seismicity of the area for many years, increasing the completeness magnitude due to waveform superposition.

## 2.2 Study area

To start the study of the Japan seismicity, first of all we established the spatial extension to take into account. We based our decision on different aspects: the magnitude of completeness ($Mc$), the configuration of the seismic networks and the distribution of the volcano-seismicity. We estimated the spatial distribution of Mc with ZMAP software (Wiemer, 2001) using a grid-based approach for the entire analyzed period 1973-2024. Specifically, the Mc was estimated using the maximum curvature method (Wiemer and Wyss, 2000), which was increased by 0.2 to account for the underestimation of completeness magnitude of this method (see Fig. 2a). The Mc of an earthquake catalog can vary over time, leading to potential underestimation if there are periods where the Mc increases for durations shorter than the catalog's entire duration. Incomplete earthquake catalogs result from two factors: Seismic Network Incompleteness (SNDI), which arises when earthquakes are difficult to detect due to a low signal-to-noise ratio (Lippiello & Petrillo, 2024), and short-term aftershock incompleteness (STAI), which occurs due to detection issues in the aftermath of large earthquakes (de Arcangelis et al., 2018). STAI is primarily caused by the masking effect of small aftershocks obscured by coda waves from previous larger ones. Given these considerations, we selected the region based on both SNDI, by examining the seismic stations depicted in Fig. 2b, and by qualitatively assessing the overall completeness magnitude. This dual approach ensures a more accurate representation of earthquake detection capabilities within the chosen region. Fig. 2a plot also shows the location of the Mt. Fuji that we did not include in our analysis. A further selection was performed on the earthquakes depth: we limited our analysis up to 50 km of depth, in order to not to consider events characterized by different geodynamic context (intermediate and deep earthquakes) in which viscoelastic effects can affect the seismicity characteristics (see also Anyfadi et al., 2023; Brondi et al., 2024; Petrillo et al., 2020, 2022; Lippiello et al. 2019b, 2021). The borders of the area run very close to the coast, except for a southern elongation related to the small Pacific islands, far from the main islands but well covered by the seismic network. In this way, we avoided the higher uncertainty on the location and magnitude associated with the events occurring offshore.

When selecting the data, we also kept in mind that it is not trivial to mix tectonic and volcanic seismicity, as the latter may have different seismic properties, leading to uncertainties in defining the classes (A and B) we are interested in. For this reason, we avoided active volcanic areas where volcanic seismicity is stronger, such as Mount Fuji and Miyake Island, located south of Mount Fuji, Miyake Island was also excluded as it is located offshore and its distance from the coast makes the estimation of earthquake location and magnitude less accurate.

Considering the characteristics of the database, we decided to choose for the analysis only the cluster with mainshock magnitude over a time-dependent threshold. First of all, NESTORE needs at least a completeness magnitude of 2 magnitude units between the Mc and Mm to evaluate the features of seismicity to be used in classification procedure (Gentili et al., 2023); in addition, in the choice of minimum Mm (Mmin) it's crucial to consider the variations in detection capabilities and location accuracy over time, the specific impacts of significant seismic events like the 2011 Tohoku earthquake and the characteristics of the catalog. The adjustment of Mmin in our study aligns with observed changes in seismicity and the evolution in earthquake detection.

- From 1973 to October 1997, we set Mmin=6. During this period, older seismic network technology may have limited the detection of smaller magnitude earthquakes. The higher Mmin value acknowledges that smaller events might not have been consistently recorded or detected, which is a common issue in earlier seismic catalogs (Nanjo et al., 2010).
- From October 1997 to March 9, 2011 (just before the Tohoku event), we set Mmin=5. This period likely corresponds to improvements in seismic detection technology and network coverage with Mc<3 in the overall analyzed area (Tamaribuchi, 2018; Tamaribuchi et al., 2021), allowing for the recording of smaller magnitude events. The decision to lower the Mmin reflects these enhanced capabilities.
- From March 9, 2011 (after and including the Tohoku event) to the end of the catalog, we again set Mmin=6. The Tohoku earthquake had a significant impact on subsequent seismic activity, including an increase in aftershock frequency and changes in stress transfer across the region (Ueda & Kato, 2023; Iidaka & Obara, 2013). This heightened seismic activity was particularly notable in areas close to the epicenter and along the Japan Trench. Further analyses have shown that the increased seismicity was not only a short-term aftermath but has had long-term implications. Areas that previously experienced low seismic activity reported higher rates post-Tohoku earthquake (Ishibe et al., 2011). For this reason, in a large area smaller clusters can not be considered unaffected by Tohoku seismicity. On the other hand, the increase in Mmin to 6 post-Tohoku might be justified since lower magnitude events could be overshadowed or not as accurately recorded due to the high volume of seismic activity. Indeed, Omi et al. (2013) found that, particularly following large earthquakes such as the 2011 Tohoku earthquake, the 50% detection rate for seismic events was achieved for magnitudes greater than 4.0. Additionally, after the 2016 Kumamoto earthquake, Zhuang et al. (2017) demonstrated the presence of missing events for magnitudes strictly greater than Mw=3. This indicates that both significant earthquakes strained the seismic networks, leading to an increase in the Mc and highlighting the challenges in detecting smaller seismic events immediately following major shocks.

# 3. Methods

NESTOREv1.0 is composed of four modules: the cluster identification module, the training module, the testing module and the Near-Real-Time (NRT) classification module. In this application, we substituted the window-based cluster identification module, we added a seismicity feature for the classification and we developed a new method for data cleaning by outlier removal.

## 3.1 Cluster identification and selection

The cluster identification procedure was one of the most challenging parts of this application to Japan: in fact, NESTOREv1.0 has a module expressly devoted to cluster identification, derived from ZMAP software (Wiemer, 2001). This uses a window based approach, in which the earthquakes belonging to the cluster are the ones inside a radius and a time duration, which are functions of the magnitude of the mainshock or of the already detected aftershocks. NESTOREv1.0 allows users to enter region-specific functions (for space and time), to adapt to different regions of the world. After several tests, we obtained that Uhrhammer (1986) laws are the most suitable for Japan seismicity but, due to the spatial proximity of different seismotectonic regions and the closeness in time of different clusters, the method fails in detecting clusters properly. For this reason, we developed a hybrid deterministic-probabilistic approach to detect earthquakes in Japan. In particular, we applied a combined cluster identification approach using the Stochastic Declustering method (Zhuang et al., 2002) based on ETAS.

Probabilistic approaches, such as the Stochastic Declustering method (Zhuang et al., 2002) based on ETAS (Ogata & Zhuang, 2006), allows us to estimate the probability that an event is spontaneous or triggered by others. The conditional intensity of the ETAS model is a summation of contributions from the background seismicity rate and from each of all previous events (triggering part)

$$\lambda(t, x, y, m) = s(m)[\mu(x, y) + \sum_{i:t_i<t} \zeta(t - t_i, x - x_i, y - y_i, m_i)] \qquad (1)$$

where $\mu(x, y)$ is the time independent background rate, s(m) is the probability density function of the Gutenberg-Richter law and $\zeta(t - t_i, x - x_i, y - y_i, m_i)$ is the triggering function which includes the Omori-Utsu law, the productivity law and the spatial distribution. We would like to underline that the standard ETAS model employs magnitude independence between seismic magnitude (Petrillo & Zhuang, 2022, 2023).

The probability that an event *j* is a background event is given by:

$$\psi_j = \mu(x_j, y_j) / [\mu(x_j, y_j) + \sum_{i:t_i<t} \zeta(t - t_i, x - x_i, y - y_i, m_i)] \quad (2)$$

conversely, the probability that an event j is triggered by a previous event i is given by:

$$\rho_{ij} = \zeta(t_j - t_i, x_j - x_i, y_j - y_i, m_i) / [\mu(x_j, y_j) + \sum_{i:t_i<t} \zeta(t - t_i, x - x_i, y - y_i, m_i)] \quad (3)$$

To employ the stochastic declustering procedure, the knowledge of the ETAS parameters is fundamental. They can be estimated through MLE direct optimization (Ogata & Zhuang, 2006), Bayesian inversion (Ross, 2021; Ross & Kolev, 2022; Molkenthin et al., 2022; Petrillo & Zhuang, 2024) or alternative methods (Petrillo & Lippiello, 2021,2023), in which the Likelihood function is difficult to calculate.

Once the parameters have been estimated, the pseudocode of the stochastic declustering algorithm is the following (the events are numbered in chronological order):

For each event *j* of the catalog:
1. estimate the probability $\rho_{ij}$ of the event j being triggered by a previous event i (i.e. *j* is an offspring of *i*) and the probability $\varphi_j$ to be background

$$\varphi_j = 1 - \sum_{i=1}^{j-1} \rho_{ij} \quad (4)$$

2. Generate a uniform random number $U_j$ in [0,1]
3. If $U_j < \varphi_j$ the *j-th* event is considered a background event
4. Otherwise, the *j-th* event is considered to be a descendant of the *n-th* event where *n* is the smallest number such that

$$U_j < \varphi_j + \sum_{i=1}^{n} \rho_{ij} = w_{nj} \quad (5)$$

By construction, the clusters catalog identified by this method is not unique, and it changes at every run, because the association of an event to a given cluster containing the event *i* depends both on the value of $w_{nj}$ and on the value of $U_j$. However, for the NESTORE application, we needed a more deterministic approach. The main idea in this case was to consider the cluster as an oriented graph, in which the events are the nodes and the links' weights are the values of $w_{nj}$. The Stochastic Declustering can be seen as a pruning procedure that cuts the links between events based on the comparison between the link weight and the value of U$_j$. The events strongly connected in the cluster, remain connected for all the runs of Stochastic Declustering method, because the links are stronger (higher $w_{nj}$) and if one link is broken, the j-*th* earthquake remains in the cluster because it is connected with not only one but several events in the clusters. For this reason, we decided to consider as belonging to the cluster only the events strongly connected to the others, in order to avoid "noise" caused by events belonging to different clusters.

The pseudocode of the cluster identification procedure consists in three steps:

1) Estimate the clusters by *k* different runs of the Stochastic Declustering procedure. Applying a trial and error approach we set *k*=10 to balance the importance of generating different clusters and the huge time consuming procedure;
2) Define one representative earthquake for each cluster; since stronger earthquakes are the ones that are more likely surrounded in time and space by clusters of seismicity, we chose the larger earthquake (the mainshock);
3) For each representative, intersect the corresponding cluster for *k*-outputs of point (1) in order to provide a unique output containing, for each mainshock, only the stable events having a high probability to belong to a given cluster;

We qualitatively assessed that the Uhrhammer (1986) space law may represent the upper limit of the extension of our clusters. We further selected in space the clusters obtained at point (3) by taking only events within a distance *d* from the o-mainshock, where d is given by the Uhrhammer (1986) law:

$$d = e^{-1.024+0.804*Mm} \ [km] \tag{6}$$

However, this choice does not affect cluster selection relevantly, because only less than 1% of the earthquakes are filtered out.

## 3.2 NESTORE training, testing and Near Real Time modules

The application of NESTORE to a dataset requires a preselection of the clusters found. To account for a plausible error in the magnitude estimate Merr=±0.1, NESTOREv1.0 discards all clusters with 0.8≤ Dm ≤1.2 where the class typology could be ambiguous. In addition, some A clusters characterized by early strong aftershocks are also discarded. Namely, NESTOREv1.0 analyzes seismicity at the end of increasing time intervals T$_i$, starting with the o-mainshock, and clusters that have already experienced an aftershock of magnitude Mm-1 do not need to be classified. The selection starts with the first time interval considered and continues for all subsequent ones: if t$_M$ is the time of the o-mainshock, t$_A$ the time of the first aftershock with Mm-1, and T$_i$ the time interval to be analyzed; if $t_A - t_M < T_i$, the cluster is not analyzed. For this reason, some A-type clusters with too early strong aftershocks are not analyzed at all, and others are only analyzed for some time intervals.

For the selected clusters, NESTORE extracts from the first hours/days after the o-mainshock a set of seismicity features, like e.g. the number of earthquakes within the cluster with magnitude M≥Mm-2, their energy, and their spatial, temporal, and magnitude distribution, and uses them during the training to differentiate between Type A and Type B clusters. More precisely, during the training phase, a one-node decision tree is employed for each feature to differentiate between the two classes.
When the decision tree converges, a threshold (*Th$_i$*) for that specific feature is found, so that most training clusters of class A have a feature value above the threshold and most clusters of class B have a feature value below the threshold. A threshold value is established for every pertinent feature, and the process is iterated for progressively extended time intervals following the mainshock, which are selected by the user and commonly span days or fractions thereof, to take into account the increase of information with time (for further details see Gentili et al., 2023). The intervals start a short time after the mainshock, typically ending a few hours/days later. The first time interval duration is set to 6 hours by default. Besides the convergence of the decision tree, the performances of the feature classifiers are evaluated through some evaluators that are the Accuracy, the Recall, the Precision, and the Informedness, that are estimated during the training by a LOO method. The first three are required to be greater than 0.5, the Informedness should be greater than 0. In addition, to take into account the dataset unbalancing, the accuracy is also required to be greater than the one of a classifier which classifies all in the more populated class (see Gentili et al., 2023). If these criteria are not satisfied in a given time period for a feature, that feature is discarded for that period.
For each feature, inside the set of time intervals in which the performances are reliable, the best performance interval is selected based on the Informedness of the feature, which is evaluated by a LOO method using the training set itself. The range between the first and the best interval duration is called the "good range". For longer time intervals, both the feature values and the threshold values are inherited from the interval with the best performance and checked with the LOO method to see whether they are still good.

It is important to remark that the thresholds are region-specific, and their numerical values depend on the seismicity itself, varying from region to region according to different seismotectonic contexts.

Following the training process, the threshold values are employed in the testing phase to derive the classification of an independent dataset: the test set. The outcomes from various features are combined to yield a conclusive classification based on Bayes' theorem (See Gentili and Di Giovambattista, 2020). This classification is compared with the actual type of the cluster (a priori known), in order to check the result of the forecasting for the different time intervals. The results obtained are shown through the Receiver Operating Characteristics (ROC) and the Precision–Recall graphs which are automatically provided in output by the NESTOREv1.0 testing module. NESTORE defines as positive the A-Type clusters and negative the B-Type ones. In particular, the ROC graph shows the True Positive Rate or Recall (TPR, normalized percentage of positive instances correctly classified as positives) vs the False Positive Rate (FPR, normalized percentage of negative instances incorrectly classified as positives). Random guessing in the ROC graph corresponds to the straight line connecting the points (0,0) and (1,1) (Egan, 1975; Sweets et al., 2000; Fawcett, 2006). The Precision-Recall graph, instead, shows Precision (the ratio between the A-type clusters correctly classified and all the positive classifications) versus the TPR (Recall and TPR coincide). An horizontal line corresponding on the Precision axis to the percentage of positive clusters in the dataset represents the random guessing. If the results are over the random guessing curve, they are considered reliable (Fawcett, 2006). The ideal performances correspond to the top left corner of the ROC graph and the top right corner of the Precision-Recall graph. If the performances estimated by the testing are considered by the user good enough, the Near Real Time (NRT) module can be applied, starting after the end of the first time interval used for the training. The NRT module makes the classification during the ongoing cluster of unknown class, and can be used, depending on the reliability of the test, for strong aftershock hazard estimation. The estimation of the time and the space for the forecasting is determined in Japan by using the Uhrhammer (1986) law.

## 3.3 NESTORE features

The features of NESTORE refer to the radiated energy, the concentration of events, the number of events and the spatial and temporal distribution. The supplementary file show the features used in detail.

The main objective of using these features is to detect changes in seismic activity, in particular increased intensity and irregularity in space, time and magnitude. This change is considered an indicator of instability within the nonlinear system associated with earthquake-generating faults (see Brondi et al., 2024 and reference therein).

An interesting feature that performed well in some previous applications of NESTORE in Italy, western Slovenia and California (Gentili and Di Giovambattista 2020, 2022; Gentili et al., 2023; Brondi et al., 2024), but not in Greece (Anyfadi et al., 2023), is the number of events in the sequence with magnitude ≥ Mm-2 (called N2), where Mm is the magnitude of the o-mainshock. For a qualitative check of its performance in Japan, we manually examined the data of the first part of the catalog we wanted to use for training NESTOREv1.0. While the feature does not seem to be very useful in class discrimination, its combination with another feature, S, which corresponds to the cumulative source area normalized to that of the o-mainshock, seems to be more efficient. Figure 3a shows the plot of S versus N2. The blue dots correspond to type B clusters and the red ones to type A clusters. Note that single feature classification is performed in NESTORE by a one node decision tree, which corresponds to a comparison with a threshold, i.e. in the figure to a line perpendicular to the feature axis. While there is no subdivision of features by class for feature N2 (no horizontal line separating the range of blue and red dots), feature S shows a more reliable, albeit noisy, performance (a vertical line separating blue and red dots can be hypothesized). However, it is possible to see a general decrease in the "red area" for increasing values of N2 (dashed line in Fig. 3a). We have formalized this qualitative result for noisy data proposing a new feature, which we call N2$_S$, where:

$$N2_s = N2 + S \cdot 110 \qquad (7)$$

Figure 3b shows the feature N2$_S$ versus N2; it seems to be more robust compared to S. These qualitative results will be tested quantitatively in session 4 on an independent dataset after data cleaning as described in the following section.

## 3.4 Data cleaning by outliers removal - REPENESE algorithm

When the dataset is strongly unbalanced like the Japan one, it is necessary to consider a preliminary assessment of the quality of the training set in order to obtain more appropriate thresholds for a clear separation of the two cluster populations in our samples.

The outlier selection algorithm to be applied to the training set should consider three important requirements:

1. Selection should only be made on the basis of the relevant features.
2. Selection should take into account the class imbalance (the elimination of a type B cluster from the dataset is less relevant than the elimination of a type A cluster, as the number of B clusters is higher).
3. The concept of distance to a cluster centroid, which is very often used in outlier detection methods such as Z-score (Rousseeuw, & Hubert, 2011) and cluster-based methods such as the Density-based Spatial Clustering of Applications with Noise DBSCAN (Ester et al 1996), should be replaced by the concept of "over" and "under" threshold, since samples with feature values that are also very far away from the centroid of the corresponding class are correctly assigned to their class if they are on the correct side with respect to the threshold.

Taking all these requirements into account, we developed a new method for detecting outliers, which we call **REPENESE** and which consists of several steps:

1. **RE** = RElevant features. To detect relevant features, we gradually increase the value of the threshold for each feature and select as relevant only those features for which we have found such a threshold that:
   1) True-positive rate>0.5;
   2) False positive rate<0.5;
   3) Precision>0.5.

2. **PE** = class imbalance PErcentage. We calculated the probability that a sample (in our case a seismicity cluster) belongs to class A or B ($P_A$ and $P_B$ respectively) as the percentage of A and B type samples:

$$P_A = \frac{Num_A}{Num_A+Num_B} \quad \text{and} \quad P_B = \frac{Num_B}{Num_A+Num_B}.$$

where $Num_A$ and $Num_B$ are the number of type-A and Type-B clusters in the training set.

3. **NE** = Neighborhood detection. For each of the relevant features estimated in point 1, we sorted the feature values of the samples. The analysis is carried on independently for each feature. For each sample of type A, the first N1 independent larger feature values are determined, and the neighborhood for the features is defined as the set of all samples that have these feature values. For type B samples, the first N1 smallest independent feature values were considered. In both cases, the number ($S_n$) of samples in the neighborhood can be greater than N1 due to the repeated values of the feature. By a trial and error approach, we chose a value of N1=5.

4. **SE**=SElection. For each sample, the number of samples N of the same class in its neighborhood was calculated and the current sample was considered a possible outlier if the following condition was met:

$$N \leq P \cdot S_n \tag{9}$$

where P corresponds to $P_A$ or $P_B$, depending on the sample class. We considered as outliers only those samples that are common for all relevant features and for all different $T_i$ for which are included in the training set, and we removed them from our cluster database before starting the training procedure.

# 4. Results

Accordingly with the previous applications of NESTOREv1.0 (Gentili and Di Giovambattista, 2020; Gentili et al., 2023; Anyfadi et al., 2023; Brondi et al., 2024), we have chosen 6 hours as the smallest time interval, for which we have 15 A and 72 B clusters. The clusters that can be used by NESTORE are shown in Figure 4. Note that, according to the NESTOREv1.0 cluster selection procedure, when an aftershock of magnitude ≥Mm-1 occurs, the cluster is classified as A and is not analyzed for longer time intervals; for this reason, the number of A-type clusters in the database usually decreases for longer time intervals.

The clusters suitable for NESTORE application are divided into two groups corresponding to a training set, on which the algorithm is trained, and a test set, to verify the training performances.

## 4.1 Training procedure and training-set

To train NESTOREv1.0, we selected the clusters from 1973 to the end of 2004. Table 1 shows the number of Type A and Type B clusters for the default time intervals of NESTOREv1.0 (every 6 hours on the first day and every day in the first week); due to early arrival of events with magnitude ≤Mm-1 for some A cluster, the number of A clusters decreases as the time interval increases.

After one day, the number of type A clusters is so low (≤4) that the REPENESE method for outlier detection (see section 3.3) cannot be used. In addition, the dataset is highly unbalanced with less than 10 % type A clusters. For this reason, we ended our analysis at 1 day.
Table 2 shows the number of available clusters after the REPENESE data cleaning, which removed 4 A- and 2 B-type clusters. This dataset was used for the training of NESTOREv1.0. Figure 5a shows the training set for a time interval of 6 hours.
We performed the training for the time intervals 0.25, 0.5, 0.75, 1 days starting with the o-mainshock. We found threshold values for 6 of the 10 features examined (the 9 default features of NESTOREv1.0 and the new feature $N2_S$ we proposed), which are listed in Table 3. Table 3 also shows the good range for every feature. Looking at the maximum of the "good range" for the features it is possible to see that for a time interval of one day, all thresholds are inherited, so no further information is extracted from the data. For this reason we stopped the classification at the previous interval: 0.75 days (equivalent to 18 hours).

The features that the training module selected for discrimination of A and B clusters, besides the already mentioned S and $N2_S$, were SLcum and SLcum2, which are the cumulative deviation of S from the long-term trend calculated with incremental and sliding windows, respectively, and QLcum and QLcum2, which are the cumulative deviation of the normalized cumulative energy from the long-term trend calculated with incremental and sliding windows, respectively.

## 4.2 Testing procedure on test-set

We tested NESTOREv1.0 on the part of the cluster database shown in Fig. 5, selecting all clusters from 2005 onwards. The test set consists of 31 clusters (4 A and 27 B - see Fig. 5b). Note that the outlier detection is not performed on the test set in order to obtain a more robust evaluation of the actual algorithm performances. Considering that the number of clusters selected for training (after outlier removal) is 50, the ratio between the total number of instances of the test set and the training set is equal to 0.62. For the type A clusters, this ratio is slightly smaller, being 0.57. These values are within the range suggested by Xu and Goodacre (2018) for an appropriate split between training and test set (50-70%) and are also compatible with the ranges chosen in previous applications of NESTORE (Gentili and Di Giovambattista, 2022; Gentili et al., 2023; Anyfadi et al, 2023; Brondi et al., 2024).

The results of the test are shown in detail in Figures 6 and 7. We have chosen as "positive" class the class "A" and as "negative" class the class "B".

In particular, Figures 6a and 6b show the performance of the features of Table 3 in the different time intervals. Time intervals are shown close to the symbol of the corresponding feature. The best performance in terms of False Positive Rate (lowest values) and Precision (highest values) were obtained by the feature $N2_S$ 6h after the mainshock (0.25 days), while SLcum2 feature obtained the best Recall at 18 hours (0.75 days) after the mainshock. Both QLCum and QLCum2 features show bad performances, close or even under the random response line. Figures 6c and 6d show the final result of NESTOREv1.0 Bayesian classification for all analyzed time periods using ROC and Precision Recall plots. Comparing upper and lower figures it is possible to see that best performances of NESTOREv1.0 were obtained six hours after the o-mainshock, where $N2_S$ outperforms S, and the NESTORE performances coincide with the ones of $N2_S$.

Figure 7a shows the forecasting results (correctly or incorrectly forecasted) in the time interval of 6 hours, while Figure 7b shows the estimated probability that a cluster is of type A (P(A)) during time; each subplot corresponds to a cluster of the test set; the yellow background color highlights the two clusters that are always incorrectly forecasted in all time intervals. Three out of four clusters of type A (75%) are forecasted correctly for all time intervals, while the percentage of correct forecasting of type B clusters decreases slightly with increasing time from 96% at 6 hours to 93% at 12 and 18 hours. Accordingly, the overall forecasting is successful in 93% of the cases 6 hours after the o-mainshock and 90% of the cases 12 and 18 hours after the o-mainshock (Fig. 7b). It is noteworthy that our test set also includes the Tōhoku sequence in 2011 (row 4, column 1 in Fig. 7b).

## 4.3 Further information based on classification

Accordingly with Brondi et al. (2024) the classification into Type-A or Type B class not only supplies some information on the stronger aftershock but, based on the statistical analysis on past clusters in the same area, can supply further information on cluster duration, the number of aftershocks with magnitude M≥Mm-2, and their cumulative seismic moment release. To obtain the seismic moment $M_0$, the conversion methods of $M_{JMA}$ to Mw, that follow a linear approach in a wide range of magnitudes, should not be applied in Japan due to the non-linear relationship between the two variables (Matsu'ura, pers. comm.). For this reason, we used the curve by Utsu et al. (1982) for average magnitude difference between $M_{JMA}$ to Mw, converted Mw into $Log(M_0)$ and expressed $Log(M_0)$ as function of $M_{JMA}$. The converted curves, together with their best fit is shown in Figure 8.

Analyzing Japan catalog from 1973 to 2023, we found that for 76% of clusters B in Japan the duration is less than half of the duration given by the window method by Uhrhammer (1986). In contrast, 67% of the A clusters in Japan have a duration greater than or equal to half of the duration determined by the window method – see Figure 9a. As for the aftershocks with M≥Mm-2, Fig. 9b shows that 80% of the B clusters have a number of less than 3, while 100% of the A clusters have at least 3 events (Figure 9b). Finally, regarding the cumulative seismic moment of the aftershocks with M≥Mm-2 normalized to the mainshock seismic moment (Figure 9c), the two types of clusters are well separated (<0.1 for B type clusters, ≥0.1 for A-type ones). This means that all A clusters have a cumulative aftershock moment with M≥Mm-2 that is greater than or equal to one-tenth of the seismic moment of the main earthquake, while this ratio is less than one-tenth for all B clusters.

## 4.4 NESTORE application in near real time

Considering the good performance of NESTOREv1.0 confirmed by the test set, we applied the NESTOREv1.0 NRT module to a new cluster. Previously, we used REPENESE to review all the data available in the training and test set and retrain the algorithm with the selected dataset. This also allowed us to compare the thresholds obtained with a larger dataset from 1973-2023 with the thresholds obtained when training with the 1973-2004 data.

In addition to the outliers found when analyzing the 1973-2004 training dataset, REPENESE correctly added the two misclassifications from the 2005-2023 test dataset (see Fig. 7b - the two clusters always misclassified outlined in yellow) to the list of outliers. It also added one outlier B before 2005, which was ignored when analyzing the smaller dataset, probably due to the larger data set and the more reliable assessment of class separation when more data were available.

Following the approach of Anyfadi et al. (2023), we trained NESTOREv1.0 with the resulting dataset between 6 and 18 hours, i.e. the range in which the software has already been tested, as we do not have a test set to validate the larger training set (see Table 4).

The obtained thresholds for N2s remained unchanged, which confirms the good stability of this feature, while the thresholds for S, SLcum and SLcum2 increased by about 25%, due to the higher number of available data used to better define the class boundaries. QLcum and QLcum2 did not converge. This is in good agreement with the poor performance of these features in Fig. 6a and 6b.

We applied the NRT to the sequence after an $M_{JMA}$ 6.6 earthquake which struck Shikoku on April 17, 2024 at 14:14 UTC and was recorded by the Japan Meteorological Agency (JMA). The epicenter was located 18 km WSW of Uwajima City, at the limit of our depth range: 48.7 km. In the following months, the Japan Meteorological Agency recorded more than 3500 aftershocks with a magnitude >0 in the region, including a magnitude 5.1 quake at 14:19 UTC, a magnitude 4.2 quake at 14:27 UTC on the same day and a magnitude 4.5 quake a month and a half later on June 1, 2024. According to the Japanese press (The Asahi Shimbun), the damage was limited: due to the earthquake-proof buildings, only 8 people were injured and a section of a national route was closed due to landslides. Since no aftershock with a magnitude greater than 5.1 was recorded, the maximum magnitude of the aftershock was less than 5.6, and the sequence evolved into a B-cluster at least until June 26 (the day we are writing this article). The sequence is still ongoing, and the forecast is issued for 197 days after the o-mainshock (the ending time for the forecasting is October 31, 2024) and a circle of 73 km around the epicenter of the mainshock, according with Uhrhammer (1986) law. Figure 10 shows the map of the sequence and its time-magnitude diagram.

We applied the NRT module of NESTOREv1.0 both with the training obtained in Section 4.1 (Table 3) and with the training on all available clusters (Table 4).

Figure 11 shows the results of NESTOREv1.0 after 6, 12 and 18 hours. This result is independent of which of the two training sets we chose (1973-2004 or 1973-2023) and it classifies the cluster as type B (the probability that it is A is estimated to be 0).

To analyze the results in more detail, we compared the classification of the individual features: Figure 12 shows the probability of being A as a bar for each of the features used; white space means $P(A)=0$, if $P(A)<0.5$ the bar is blue, otherwise red; if the feature is not available, this information is given instead of the bar. For the features, we consider A classification of $P(A)\geq0.5$. The smaller training set provided a final NESTOREv1.0 classification of the cluster as B (see Fig. 12), a B-Type classification for all features 6 hours after the o-mainshock, but an A-Type classification (red bars in Fig. 15a) for longer time intervals due for QLcum and QLcum2 features. Although this result did not affect the NRT classification of this cluster, it could in principle lead to uncertain classifications of NRT for other clusters. When training NESTOREv1.0 with a larger training set, QLcum and QLcum2 features were evaluated as less reliable and the corresponding classifiers were removed from those used for the final classification (see Fig. 12b). Accordingly with the results of section 4.3, if the class is B, we performed the following estimates:

1. Less than 3 events with magnitude ≥4.6 with a probability of 80%
2. Cluster duration <99 days with a probability of the 76%. Note that for the duration the forecast refers to cluster' events with magnitude ≥Mm-2=4.6
3. Ratio between cumulative moment of the aftershocks and the moment of the mainshock <0.1 for all B clusters

At June 26, 2024, 70 days after the mainshock, JMA recorded only one aftershock with magnitude ≥4.6, 5 minutes after the mainshock, so, by now, conditions 1 and 2 are satisfied. The ratio between the cumulative moment of the aftershocks and the mainshock moment is 0.0065. These data are in agreement with the results of section 4.3 but the validation will be on October 31, 2024.

## 5. Discussion and conclusions

The issue of forecasting and/or estimating the probability of occurrence of strong earthquakes is of paramount importance in Japan. Several research papers have addressed this issue from different angles at regional (e.g. Hirose et al., 2011) and local levels (e.g. Parsons et al., 2012; Fukushima et al., 2023) after strong events or in high-hazard areas, often relying on statistical magnitude-frequency relationships (e.g. the Gutenberg-Richter law). The Japanese government has also systematically addressed the prevention of earthquake disasters and created special structures to deal with the hazard (e.g. the HERP, https://www.jishin.go.jp/main/index-e.html ). In addition, the Japanese Meteorological Agency has promoted an advanced early warning system since 2007 (Hoshiba et al., 2008), which was also improved after the magnitude 9.0 Tōhoku earthquake in 2011 (Kodera, 2021). These examples illustrate the attention and efforts to mitigate seismic risk, and in this regard, the main objective of our research is to support them by providing a tool to calculate the probability that an earthquake could be followed by strong aftershocks.

The NESTORE machine learning algorithm contributes to the forecasting of strong earthquakes during the occurring seismic clusters by providing a binary classification for a given earthquake cluster: Type A if strong aftershocks are expected and Type B if not. The classification is based on a set of seismological features extracted from the first hours/months of seismicity. NESTORE has already been successfully used in various seismological regions: Italy, Slovenia, California and Greece (Gentili and Di Giovambattista, 2017, 2020, 2022; Anyfadi et al., 2023; Brondi et al., 2024). The application to Japan represents a major challenge for the algorithm, as the seismicity there is very high and consequently a huge amount of data has to be processed. These earthquakes are the result of the complex seismotectonic context of this region, which is affected by the interaction of three tectonic plates and where very strong earthquakes frequently occur, both in the crust and at depth, triggered by volcanic and tectonic processes.

In this study we tested NESTORE on a set of seismic clusters that occurred in the about 20 years, whose type is already known, in order to compare the forecasted classification with the real one. This allowed for getting feedback on the forecasting accuracy of the code and its potential employment also in near real time application, a module already implemented (Gentili et al., 2023). However, the testing procedure has been just the last step of a more comprehensive statistical analysis, which comprised (1) an evaluation of the minimum magnitude of the mainshock to be taken into account (2) the development of a new cluster identification method based on Stocasting declustering by ETAS method (Zhuang et al., 2002), (3) the proposal of a new feature of seismicity expressly suited for discrimination of A and B clusters in Japan, and (4) an innovative outlier detection method. We discuss in the following the key points of this work.

1. The minimum magnitude of the mainshock to be considered is related to the completeness magnitude, because NESTOREv1.0 needs a completeness magnitude at least 2 grades under the one of the o-mainshock. The completeness magnitude changed in time and space in Japan, due to different reasons not only related to the number and the position of seismic stations but also to the management of the catalog, which changed in time, and to the Tohoku earthquake sequence, which increased the number of occurred earthquakes but increased also the completeness magnitude, due to waveform superposition. Another aspect of the Japan seismicity after Tohoku 2011 is that the long duration of the sequence (about 12 years according to Urhammer (1986) law) influenced the small seismicity until recent years, making it difficult to analyze independent clusters in a large area. In order to handle these variations, we selected a region of analysis, over a given depth and inside a line closed to japan costs, and three periods in which we selected a different minimum o-mainshock magnitude.

2. Cluster identification is a crucial process for assessing seismic hazards, improving risk models and enhancing earthquake forecasting. Several methods have been proposed for declustering seismic catalogs to identify background and triggered seismicity. Most simple methods remove earthquakes within a space-time window around a mainshock, with variations primarily in the choice of window sizes and detailed procedures (Gardner and Knopoff, 1974; Kellis-Borok and Kossobokov, 1986). Generally larger mainshocks require larger windows. An alternative approach is the link method which relies on space-time distances between events (Reasenberg 1985; Frohlich and Davis 1990; Davis and Frohlich 1991). All conventional methods involve arbitrary parameters for defining aftershock window sizes or link distances, leading to different declustered catalogs and varying estimates of background seismicity. This parameter dependency often results in subjective choices influenced by the user's impression of the data. Additionally, the concept of what constitutes an aftershock is not uniquely defined, prompting the development of numerous declustering algorithms. To address these issues, stochastic models have been suggested to objectively quantify observations, assigning a probability to each event of being a background or cluster event (Musmeci and Vere-Jones, 1992; Ogata, 1998; Zhuang et al., 2002). Stochastic models aim to provide a more objective and consistent method for estimating background seismicity, avoiding the subjective biases of conventional approaches. On the other hand, the limitation of stochastic approaches is that they do not provide a deterministic classification but only a probabilistic one. This can lead to the risk of including or excluding events that do not belong to a specific class of events (clustered or background event). To address this issue, in our study, we employed a combined approach that leverages the strengths of both stochastic and window-based methods, aiming to achieve the most realistic classification of seismic clusters possible.

3. Regarding the seismological features, we introduced a new feature ($N2_S$) based on N2 (the number of aftershocks with M≥Mm-2) and S (the cumulative source area of the aftershocks); the new feature, we developed empirically, provided good performances, showing high Accuracy and Precision in the overall forecasting of the clusters of the test-set. Its performances are higher than that of the feature N2 which, on the other hand, has been shown to be very informative in previous applications of NESTORE in Italy, Western Slovenia and California (Brondi et al., 2024; Gentili and Di Giovambattista, 2022), and S, which supplied good results in Italy, Western Slovenia and Greece (Brondi et al., 2024; Anyfadi et al., 2023). We have observed that this is related, for high values of S, to a decrease of the values of S for A clusters as N2 increases (see the dashed line in Fig. 3 - note that the opposite trend for small values of S has been recognized as due to outliers). We try to interpret this result. When an earthquake occurs, it can trigger aftershocks due to the redistribution of stress along faults. As aftershocks continue to happen, they can propagate both along the fault line and in adjacent areas. We hypothesize that the characteristic trend for type-A clusters in Japan may be due to faults structure and heterogeneous geology; the main earthquakes may rupture a portion of the fault, leaving smaller areas still under stress with specific geological features, such as high levels of strain accumulation or weak rock formations. The aftershocks may be concentrated in areas, generating small but frequent earthquakes, until a larger area breaks, when the strongest aftershock happens. Note that it is an hypothesis and its validation, which requires an extended analysis of the Japanese faults and cluster evolution is behind the scope of this paper.

4. The last innovation of this paper is the method we developed to identify outliers. It was necessary to develop it due to the large unbalancing of the dataset, where the type-A clusters are only the 20% of the analyzed ones and, as it often happens in cluster analysis applications, the total number of available clusters is small (56 clusters). Due to these characteristics of the dataset, a limited number of outliers, that is very common in real data, can affect the classifier capability to correctly estimate the class boundaries in the feature space. The standard outliers detection methods are inapplicable due to the specificity of our data: most features depend on the magnitude and number of aftershocks, and clusters with a lower number and magnitude of aftershocks are the most common; for this reason, the distribution of the two classes is generally right-skewed; in addition, while the feature values of the type B clusters should lie between the value 0 (if there is no aftershock) and the threshold, there is no upper limit for the feature values of the type A clusters, as all feature values above the threshold are well classified. The more common methods for detecting outliers are based on the hypothesis of a Gaussian distribution of instances of each class; a typical example is the Z-score (Rousseeuw & Hubert, 2011), where any instance that deviates further than, say, 3 standard deviations from the mean is considered an outlier. Other methods use percentiles, ranging from those that use the interquartile range (IQR - Rousseeuw & Hubert, 2011), which assumes a symmetrical distribution so that all data outside, say, the 5% and 95% percentiles are considered outliers, to more sophisticated methods where the upper and lower percentiles may not be symmetrical. All of these methods are not well suited for unknown skewed distributions. An unsupervised clustering method in which the model is learned based on the organization of the training samples without labeling information, DBSCAN (Ester et al., 1996), could be used to detect outliers in this case by supplying the A and B class data separately. However, it is necessary to modify the standard approach of this method, in which all instances within a certain predefined distance are considered as "neighbors" of an instance to be checked. Due to the different distribution of feature values and the isolated instances with high values belonging to class A, it was necessary to establish a new strategy for defining the neighborhood, i.e. an "oriented" neighbor whose elements are all smaller or equal or larger or equal to the analyzed instance depending if the instance belongs to the B or A class. In a way, one can say that REPENESE is inspired by the DBSCAN method, but optimized for skewed distributions, and we hope that this algorithm can be applied also to other similar problems related to different fields of study, like biology, finance or medicine.

Applying all these improvements to the NESTOREv1.0 algorithm led to encouraging results in the test set: retrospective forecasting using data from 2005 allowed us to correctly forecast 75% of A clusters and 96% of B clusters, achieving a precision of 0.75 and an accuracy of 0.94, 6 hours after the occurrence of the o-mainshock. We had the important opportunity to test the potential of the software with a cluster consisting of more than 3000 events with M>4 and including the Tōhoku major event of March 11, 2011 with Mw=9.1, one of the strongest earthquakes in the instrumental record worldwide. The Tōhoku seismic sequence, which began with a magnitude Mw=7.3 earthquake on March 9, 2011, has a $P(A)=1$ for all time intervals considered and is correctly forecasted as a type A cluster.

We also applied the algorithm to the ongoing seismic sequence that started on April 17, 2024 in Shikoku with a 6.6 magnitude o-mainshock, and the forecasting was a B-type cluster. These results will be validated on October 31, 2024 with the end of the forecasting period, but at this stage (June 26, 2024) the results are encouraging as the highest recorded aftershock corresponds to magnitude 5.1. This is the first application of the NRT module in Japan. It should be noted that in Japan, due to the enormous seismicity and complexity of the region, it can happen that during a sequence another sequence starts that is so close that it is within the sequence radius estimated by Uhrhammer's method; so far this is not the case for the 2024 Shikoku sequence. Since the system is designed to work in near real-time, the operator can exclude the earthquakes belonging to another sequence from the analysis in this case. To avoid non-automatic, possibly arbitrary evaluations, newer versions of the code will include a fast version of the cluster identification method from Section 3.1.

## 6. Author contributions

**Stefania Gentili:** Conceptualization, Methodology, Software, Validation, Formal analysis, Investigation, Resources, Data Curation, Writing - Original Draft, Writing - Review & Editing, Visualization, Supervision, Project administration, Funding acquisition; **Giuseppe Davide Chiappetta:** Software, Validation, Formal analysis, Investigation, Data Curation, Writing - Original Draft, Writing - Review & Editing, Visualization; **Giuseppe Petrillo:** Investigation, Data Curation, Writing - Review & Editing; **Piero Brondi:** Software, Investigation, Writing - Review & Editing, Visualization; **Jiancang Zhuang:** Methodology, Software, Resources, Data Curation, Writing - Review & Editing, Supervision, Project administration, Funding acquisition.

## 7. Acknowledgments

Funded by a grant from the Italian Ministry of Foreign Affairs and International Cooperation and Co-funded within the RETURN Extended Partnership and received funding from the European Union Next-GenerationEU (National Recovery and Resilience Plan - NRRP, Mission 4, Component 2, Investment 1.3 – D.D. 1243 2/8/2022, PE0000005) and by the NEar real-tiME results of Physical and StatIstical Seismology for earthquakes observations, modeling and forecasting (NEMESIS) Project (INGV). Map figures have been realized using QGIS (http://www.qgis.org), while other analyses have been performed using ZMAP (http://www.seismo.ethz.ch/en/research-and-teaching/products-software/software/ZMAP). For Figure 11 we used the free online application Plotdigitizer (https://plotdigitizer.com/app) for data extraction.

The NESTOREv1.0 toolbox is available for free download from GitHub at the address https://github.com/StefaniaGentili/NESTORE and its reproducibility package is available on Zenodo https://zenodo.org/account/settings/github/repository/StefaniaGentili/NESTORE.

We wish to thank Rita Di Giovambattista for her useful suggestions on NESTORE and Ritsuko Matsu'ura for the useful information on $M_{JMA}$ and $M_0$ scales.

Gentili, S., Brondi, P., Di Giovambattista, R., 2023. NESTOREv1.0: A MATLAB Package for Strong Forthcoming Earthquake Forecasting. Seismol. Res. Lett. 94 (4), 2003-2013. https://doi.org/10.1785/0220220327

Gulia, L., Wiemer, S., Vannucci, G. (2020). Prospective evaluation of the foreshock traffic light system in Ridgecrest and implications for aftershock hazard assessment. Seismol. Res. Lett. https://doi.org/10.1785/0220190.

Hirose, F., Maeda, K., 2011. Earthquake forecast models for inland Japan based on the G-R law and the modified G-R law. Earth, Planets and Space 63, 239–260. https://doi.org/10.5047/eps.2010.10.002

Hoshiba, M., Kamigaichi, O., Saito, M., Tsukada, S. y., Hamada, N., 2008. Earthquake Early Warning Starts Nationwide in Japan. Eos, Trans. Am. Geoph. Un. 89 (8), 73–74. https://doi.org/10.1029/2008EO080001.

Iidaka, T., Obara, K., 2013. Shear-wave splitting in a region with newly-activated seismicity after the 2011 Tohoku earthquake. Earth, Planets and Space 65, 1059–1064 https://doi.org/10.5047/eps.2013.02.003

Ishibe, T., Shimazaki, K., Satake, K. Tsuruoka, H., 2011. Change in seismicity beneath the Tokyo metropolitan area due to the 2011 off the Pacific coast of Tohoku Earthquake. Earth, Planets and Space 63. https://doi.org/10.5047/eps.2011.06.001

Japan Meteorological Agency (2024) The seismological bulletin of Japan. https://www.data.jma.go.jp/svd/eqev/data/bulletin/index_e.html.

Kawasaki, S., Matsusaki, S., Fukushima, Y., 2008. A relation between Mjma and seismic moment (determined from dense broad band seismograph network). AGU Fall Meeting Abstracts, 2008. https://ui.adsabs.harvard.edu/abs/2008AGUFM.S13C1827K

Kaizer, J., Kontuľ, I., Povinec, P. P., 2023. Impact of the Fukushima Accident on 3H and 14C Environmental Levels: A Review of Ten Years of Investigation. Molecules 28(6). https://doi.org/10.3390/molecules28062548

Keilis-Borok, V. I., Kossobokov, V. G., 1986. Time of increased probability for the great earthquakes of the world. Comput. Seismol 19, 48-58.

Kodera, Y., Hayashimoto, N., Tamaribuchi, K., Noguchi, K., Moriwaki, K., Takahashi, R., Morimoto, M., Okamoto, K., Hoshiba, M. 2021. Developments of the Nationwide Earthquake Early Warning System in Japan After the 2011 Mw 9.0 Tohoku-Oki Earthquake. Front. Earth Sci. 9. https://doi.org/10.3389/feart.2021.726045.

Lay, T., 2018. A review of the rupture characteristics of the 2011 Tohoku-oki Mw 9.1 earthquake, Tectonophysics 733, 4-36. https://doi.org/10.1016/j.tecto.2017.09.022.
28

Wiemer, S., Wyss, M., 2000. Minimum Magnitude of Completeness in Earthquake Catalogs: Examples from Alaska, the Western United States, and Japan. Bulletin of the Seismological Society of America, 90 (4), 859-869. https://doi.org/10.1785/0119990114

Wiemer, S., 2001. A software package to analyze seismicity: ZMAP. Seismological Research Letters 72 (3), 373-382. https://doi.org/10.1785/gssrl.72.3.373

Wu, J., Suppe, J., Lu, R., Kanda, R., 2016. Philippine Sea and East Asian plate tectonics since 52 Ma constrained by new subducted slab reconstruction methods. J. Geophys. Res. Solid Earth 121 (6), 4670–4741. https://doi.org/10.1002/2016JB012923.

Xu, Y., Goodacre, R., 2018. On Splitting Training and Validation Set: A Comparative Study of Cross-Validation, Bootstrap and Systematic Sampling for Estimating the Generalization Performance of Supervised Learning. J. Anal. Test 2, 249–262. https://doi.org/10.1007/s41664-018-0068-2

Zhuang, J., Ogata, Y., Vere-Jones, D., 2002. Stochastic Declustering of Space-Time Earthquake Occurrences, Journal of the American Statistical Association 97 (458), 369-380. https://doi.org/10.1198/016214502760046925.

Zhuang, J., Ogata, Y., 2006. Properties of the probability distribution associated with the largest event in an earthquake cluster and their implications to foreshocks. Physical Review E 73 (4). https://doi.org/10.1103/PhysRevE.73.046134

Zhuang, J., Ogata, Wang, T., 2017. Data completeness of the Kumamoto earthquake sequence in the JMA catalog and its influence on the estimation of the ETAS parameters. Earth Planets and Space 69. https://doi.org/10.1186/s40623-017-0614-6
## 9. Figure captions

Fig. 1: Plate boundaries representation. Data retrieved from Van Horne et al. (2017).

Fig. 2: (a) Mc map computed on the catalog from 1973 to 2024, the magenta circle shows the location of the Mt. Fuji while the red line is the study area we selected . (b) Map of the seismic stations in Japan showing the JMA and the F-NET networks (data from https://www.data.jma.go.jp/svd/eqev/data/bulletin/catalog/appendix/appendix_e.html).

Fig. 3: Plot of some of the NESTORE features: S vs N2 and N2S vs N2. Red dots: Type A clusters, Blue dots: type B clusters dashed lines: possible subdivision between classes.

Fig. 4: Available clusters for the analysis before the data cleaning. Circles: epicenters of the o-mainshocks, the color shows the clusters' classification.

Fig. 5: NESTOREv1.0 dataset for Japan: (a) training set after outliers removal (b) test set. Circles: epicenters of the o-mainshocks, the color shows the clusters' classification.

Fig. 6: Results of the testing procedure showing the performances of the features and of the overall Bayesian classification in terms of ROC (a, c) and Precision-Recall (b, d). Numbers show the time period (in days) of the corresponding symbol.

Fig. 7: (a) Results of the forecasting. All these plots refer to Ti=6h. (b) Probability of being an A-type cluster vs time.

Fig. 8: Red line: Log(M0) as function MJMA converted by Utsu's 1982 paper. Dashed line: quadratic fit.

Fig. 9 (a) Distribution of the cluster duration normalized to the time window width given by the window method (Uhrhammer 1986), (b) distribution of the number of aftershocks with M≥Mm-2, and (c) the corresponding cumulative seismic moment normalized to that of the FSE for the Japanese catalog

Fig. 10: Shikoku sequence from April 17 to June 26, 2024 (a) map of events epicenters; the red line corresponds to the limits in area accordingly with Uhrhammer's law (b) time magnitude plot; the dashed line corresponds to Mm-2.

Fig. 11: Probability that the Shikoku cluster is a type A estimated by NESTOREv1.0 at 6,12,18 hours after the o-mainshock. The forecasting is performed for a region with radius 73 km around the o-mainshock and for a time of 197 days after the mainshock.

Fig. 12: NESTOREv1.0 single feature classification at different time intervals using the two different training sets (a) Training set 1973-2004 (b) training set 1973-2023.

## 10. Table captions

Table 1: Number of A and B clusters and percentage of A clusters in the different time intervals, before removal of outliers.

Table 2. Number of A- and B-type clusters and percentage of A-type clusters through the different time intervals in the final training set.

Table 3. Thresholds of the features and their good interval computed using the training set of clusters during the years 1973-2004 after outliers removal.

Table 4. Thresholds of the features and their good interval computed using the training set of clusters during the years 1973-2023 after outliers removal.

## 11. Figures

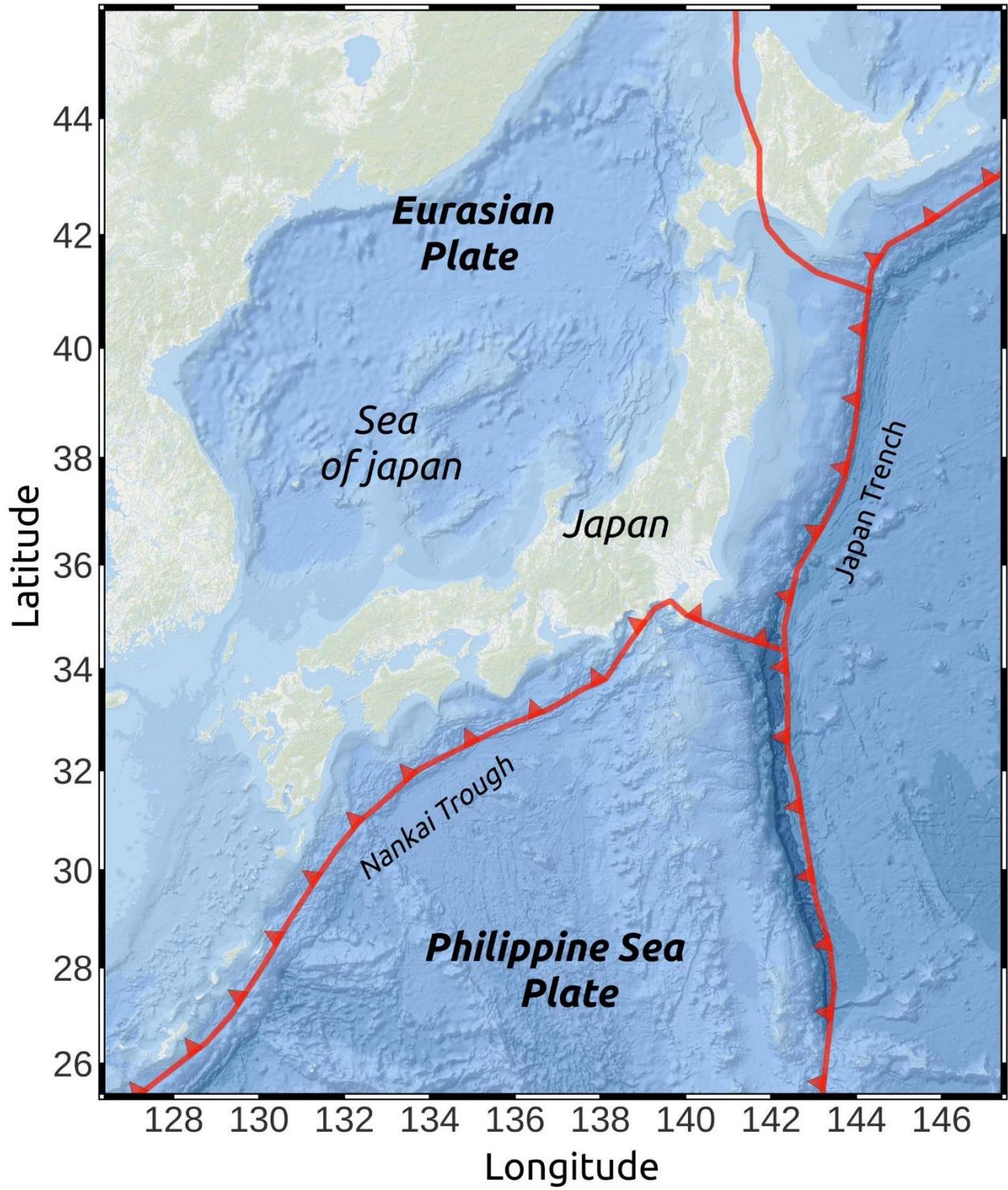

Fig1

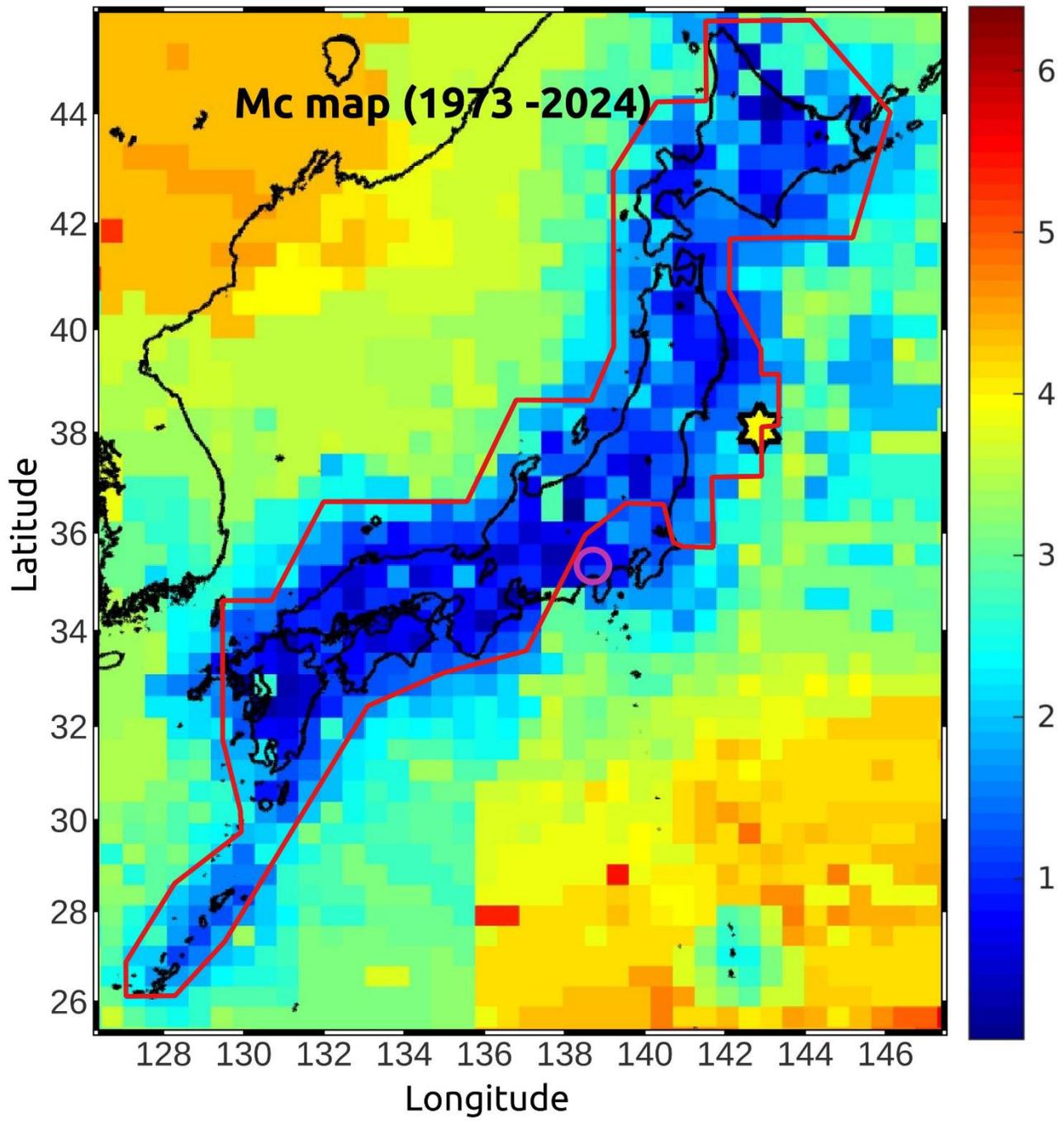

Fig2a

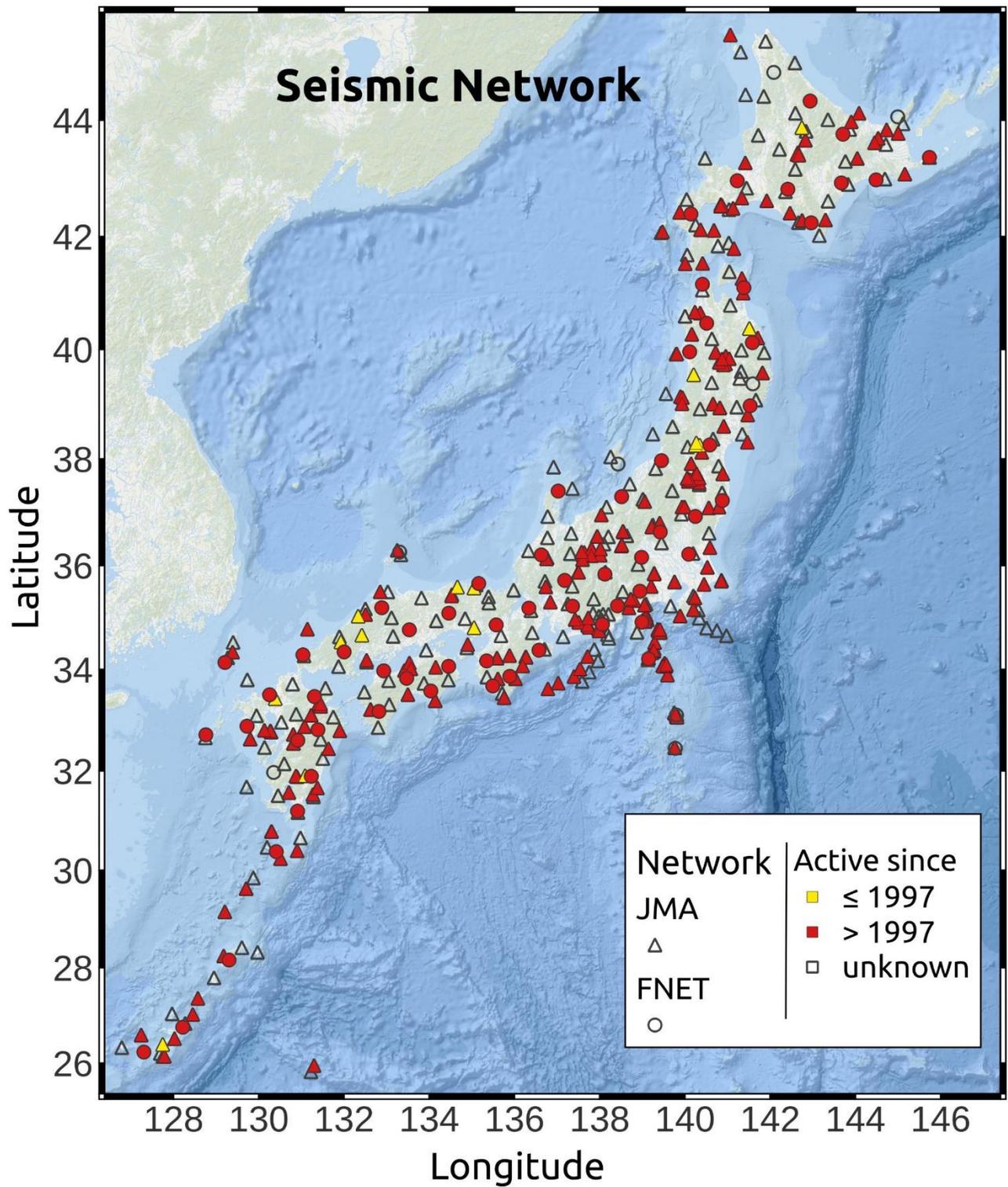

Fig2b

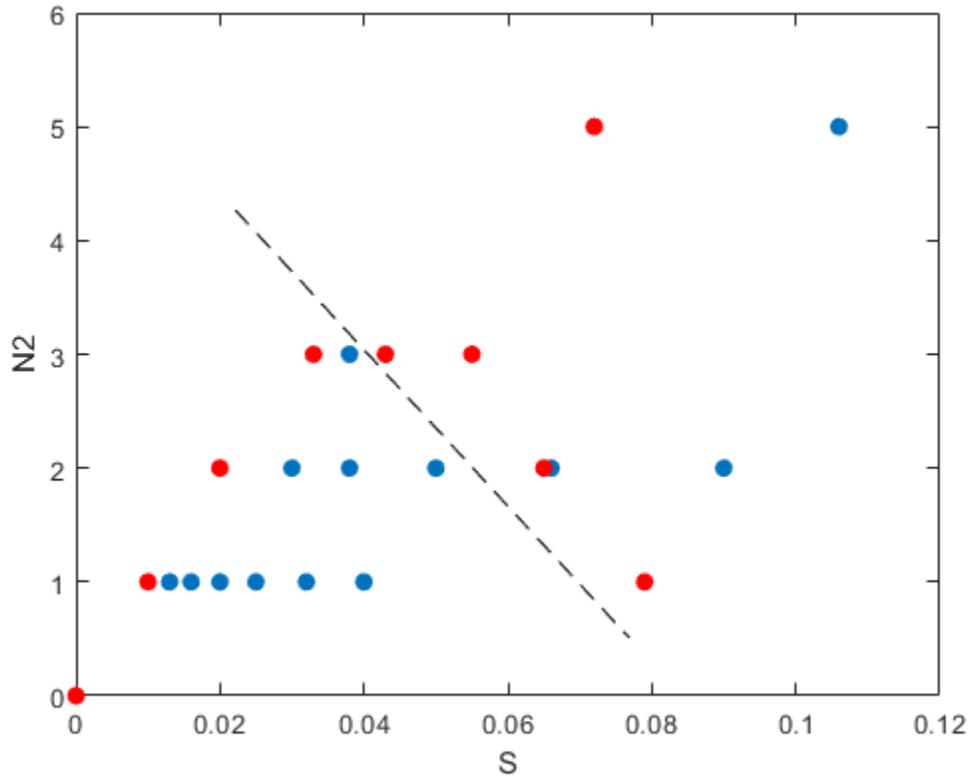

Fig 3a

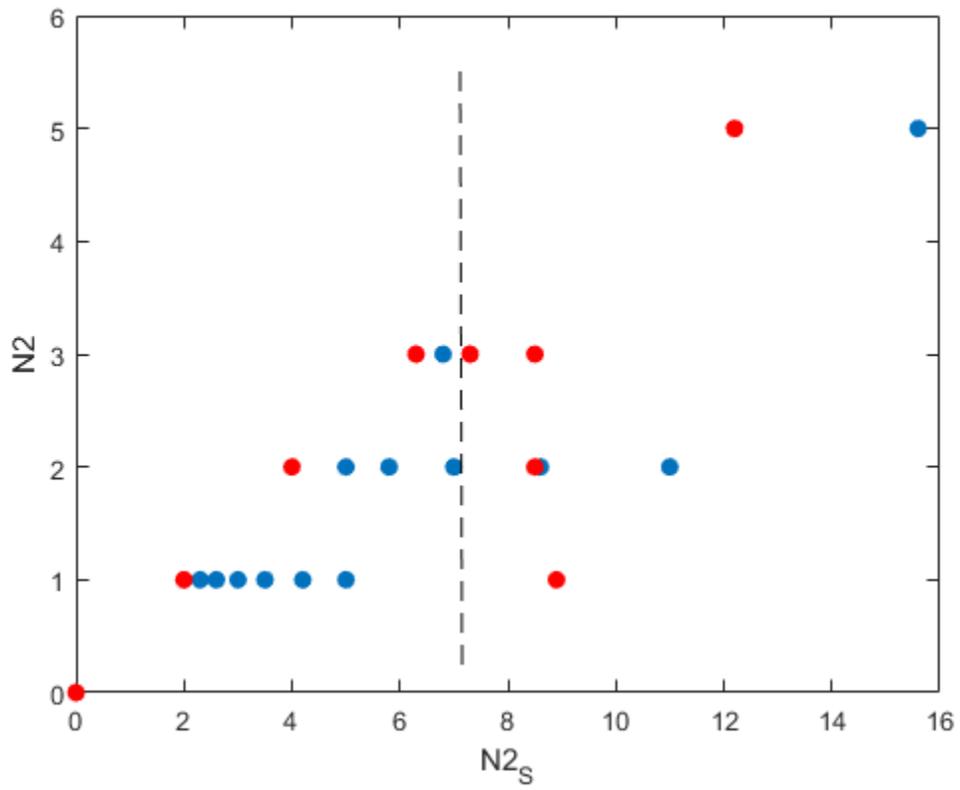

Fig 3b

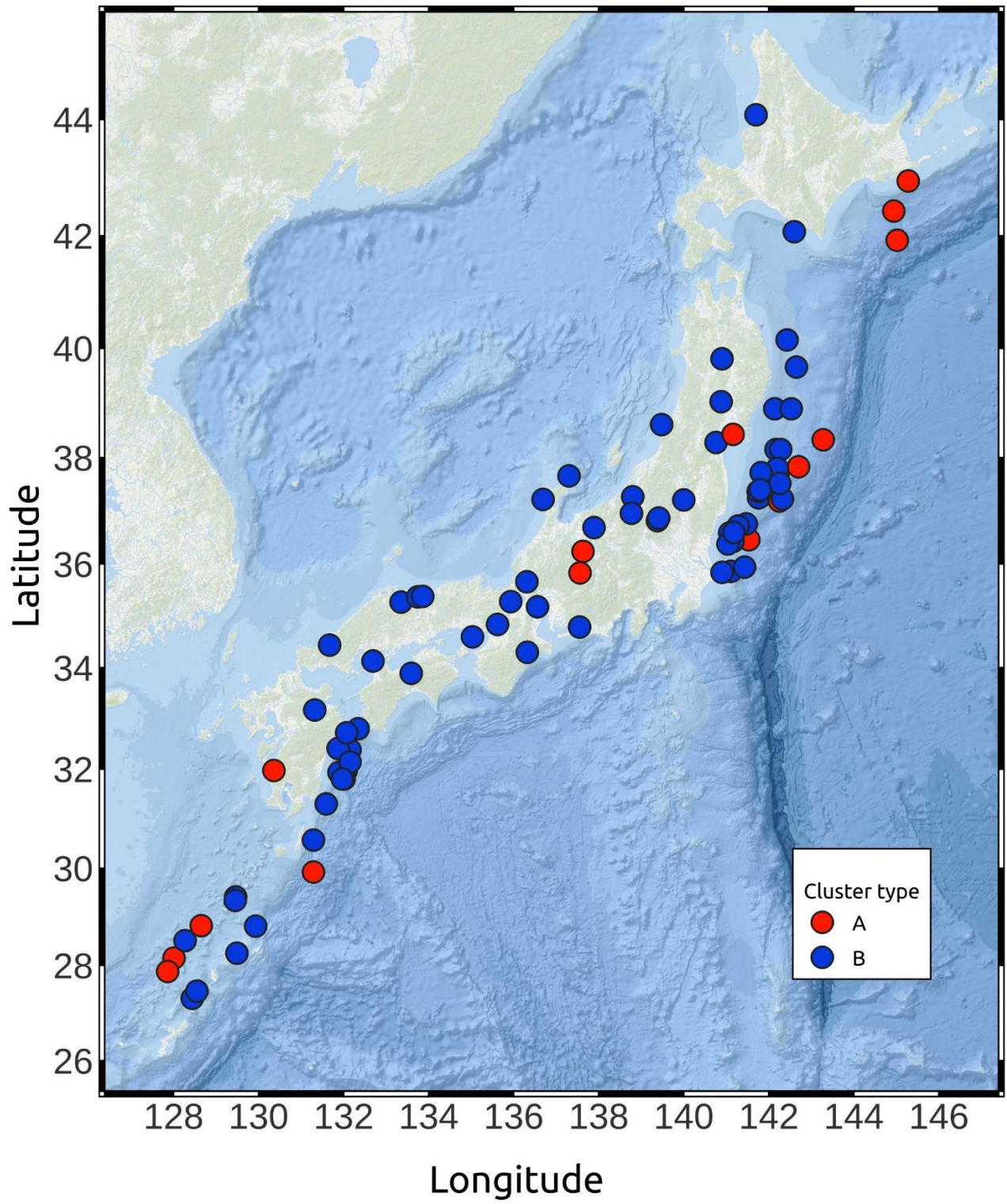

Fig 4

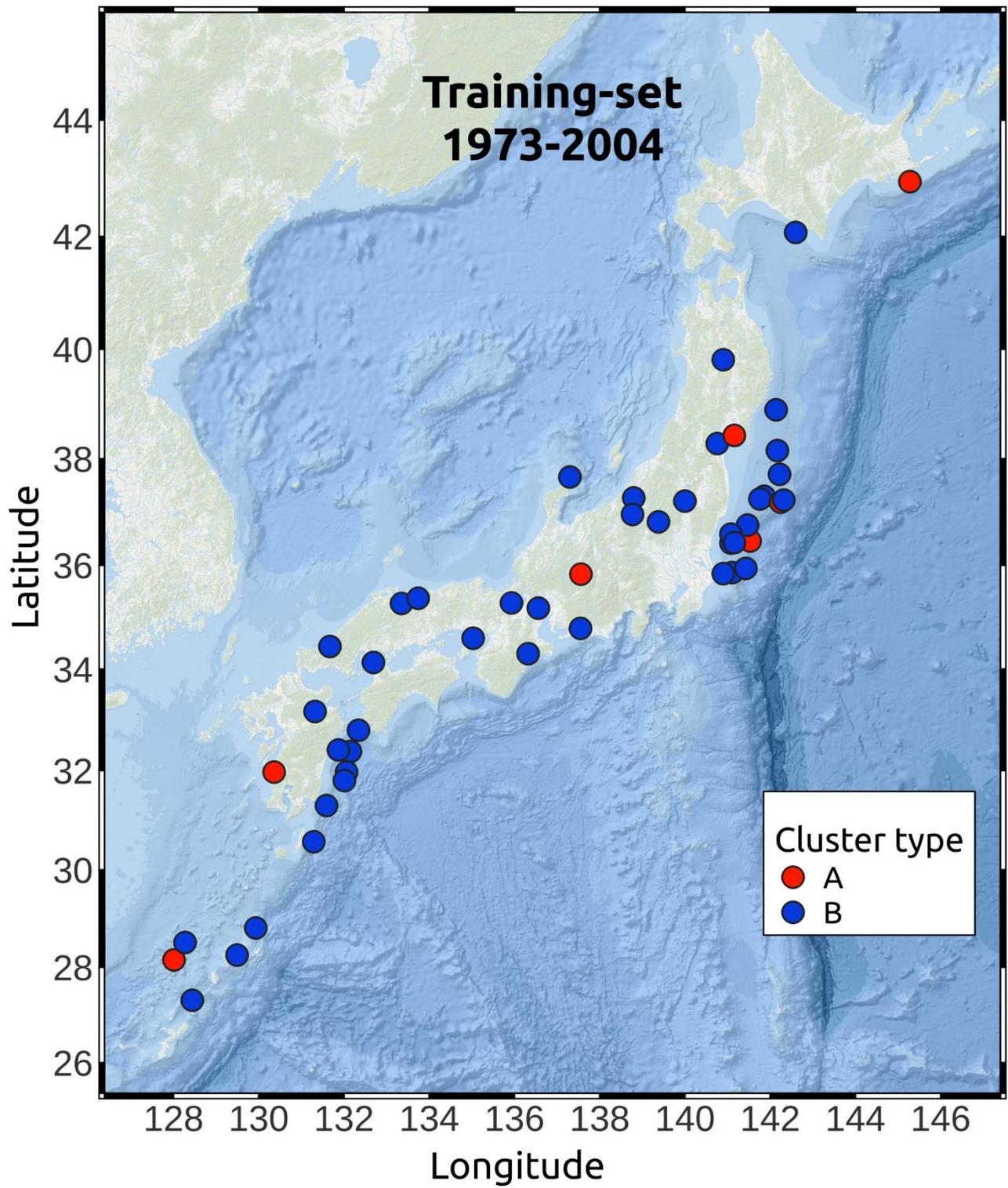

Fig 5a

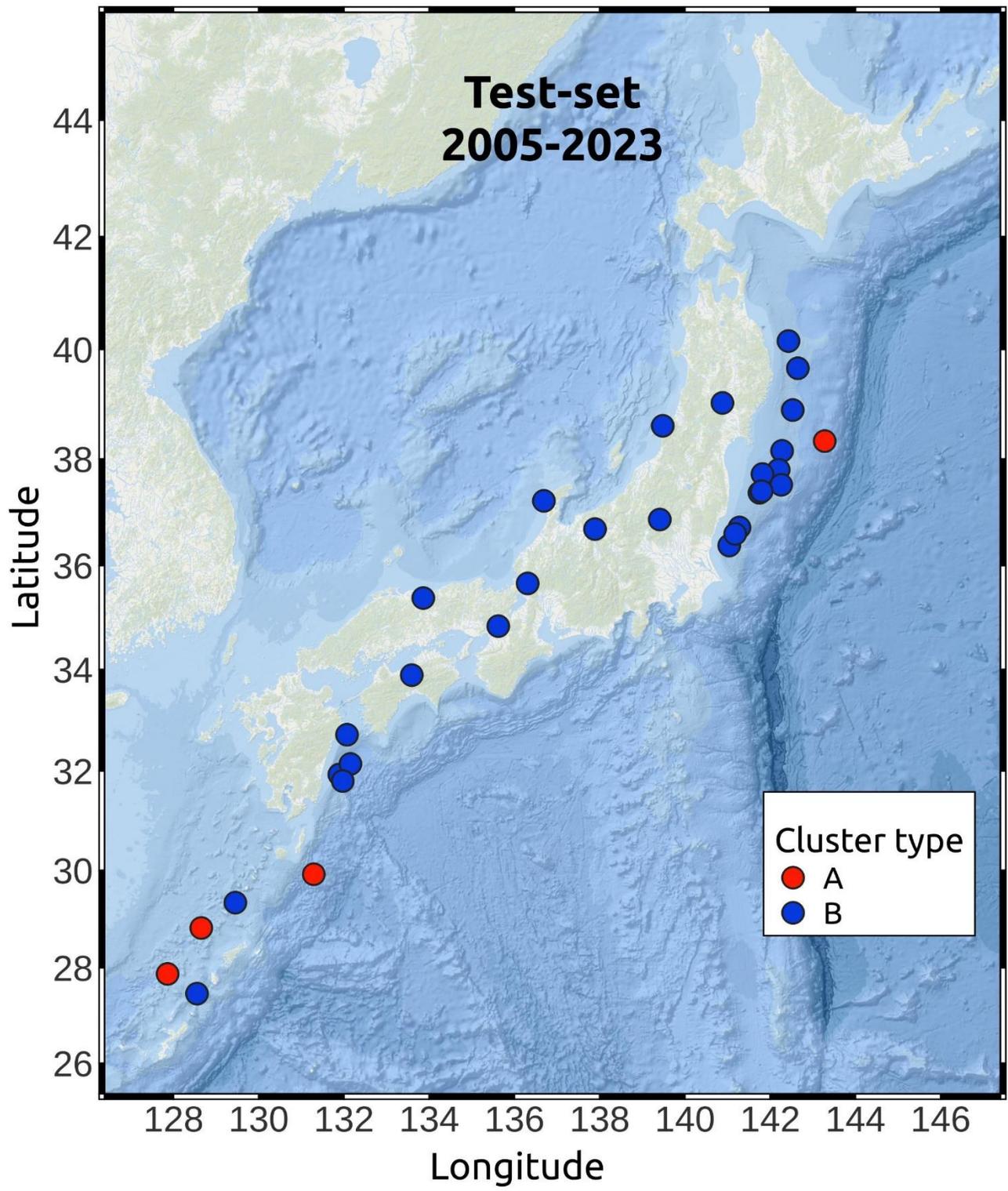

Fig5b

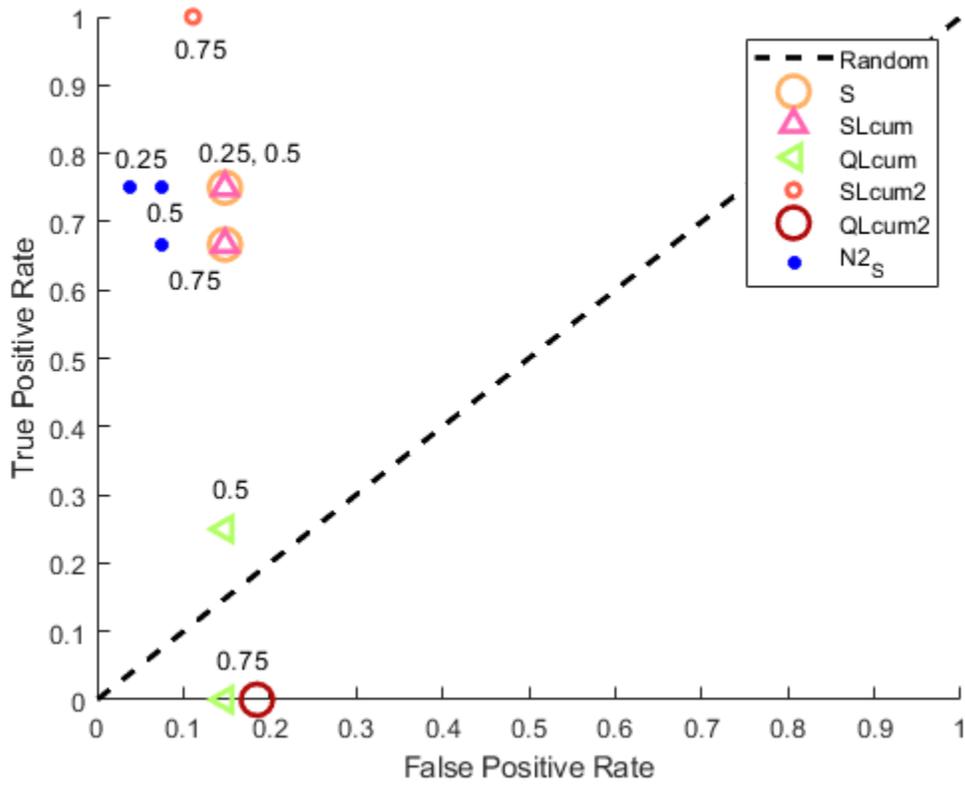

Fig6a

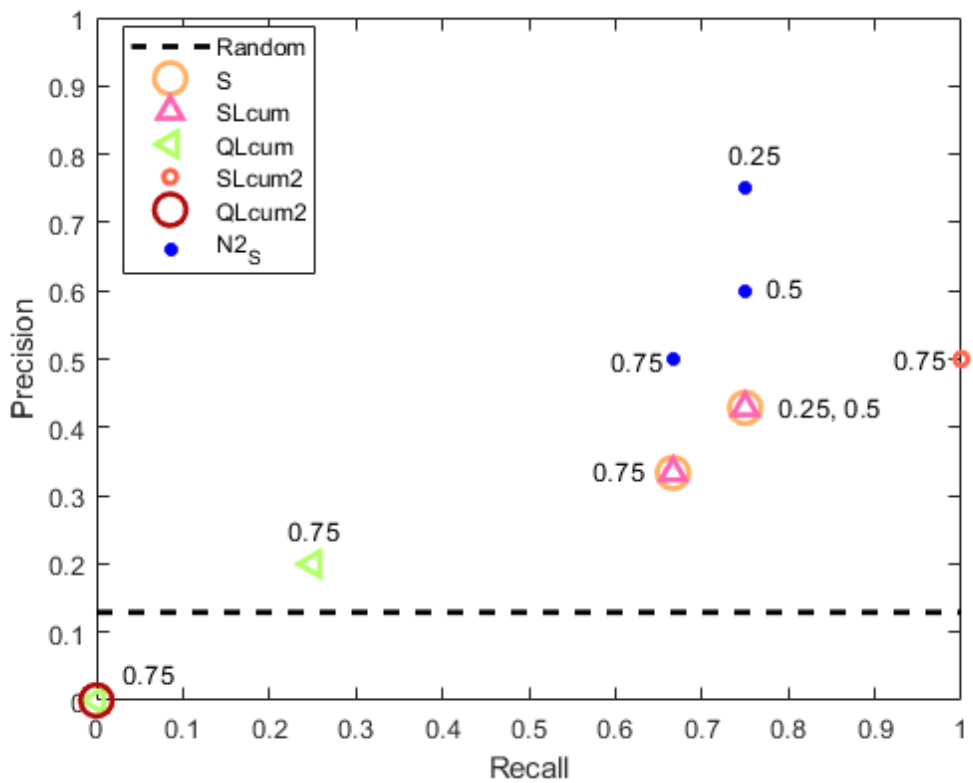

Fig 6b

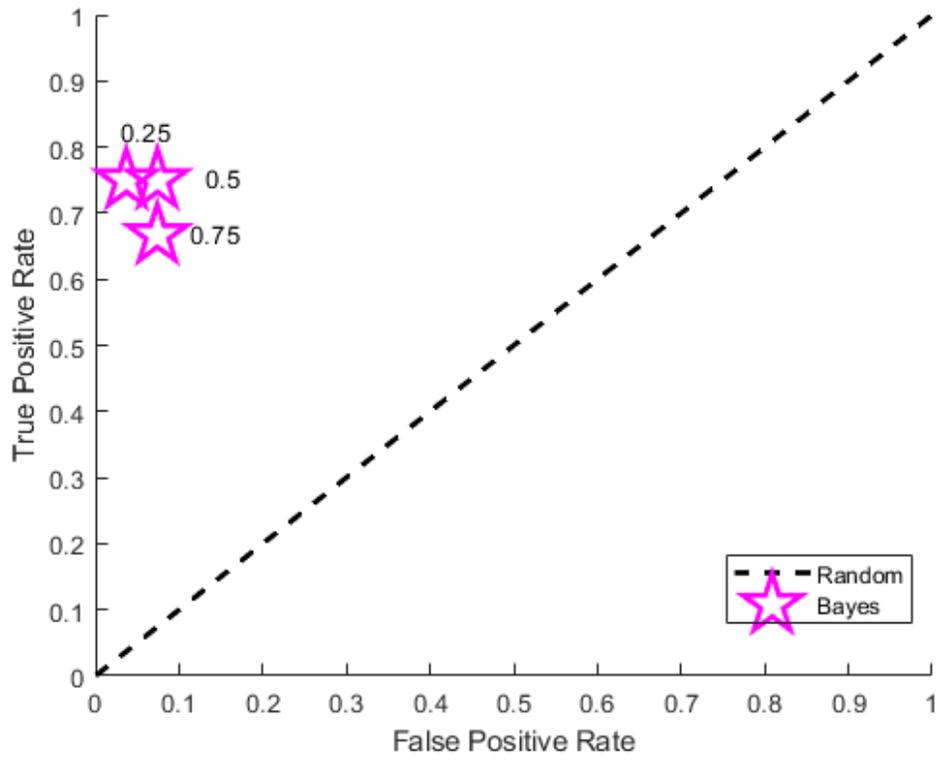

Fig 6c

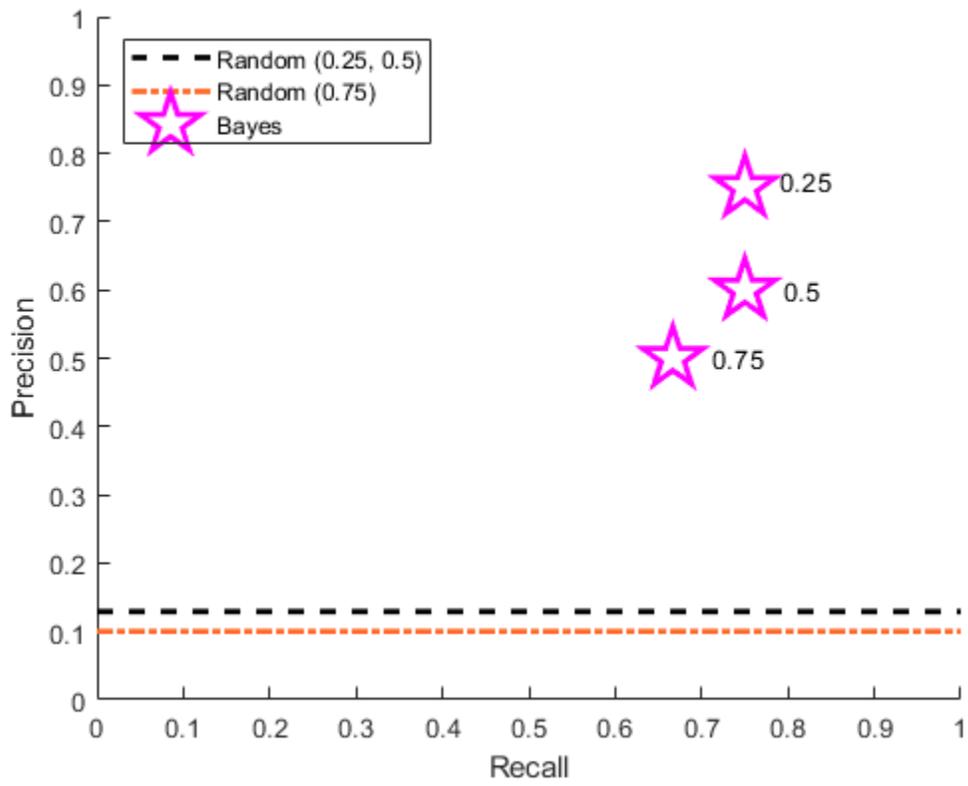

Fig 6d

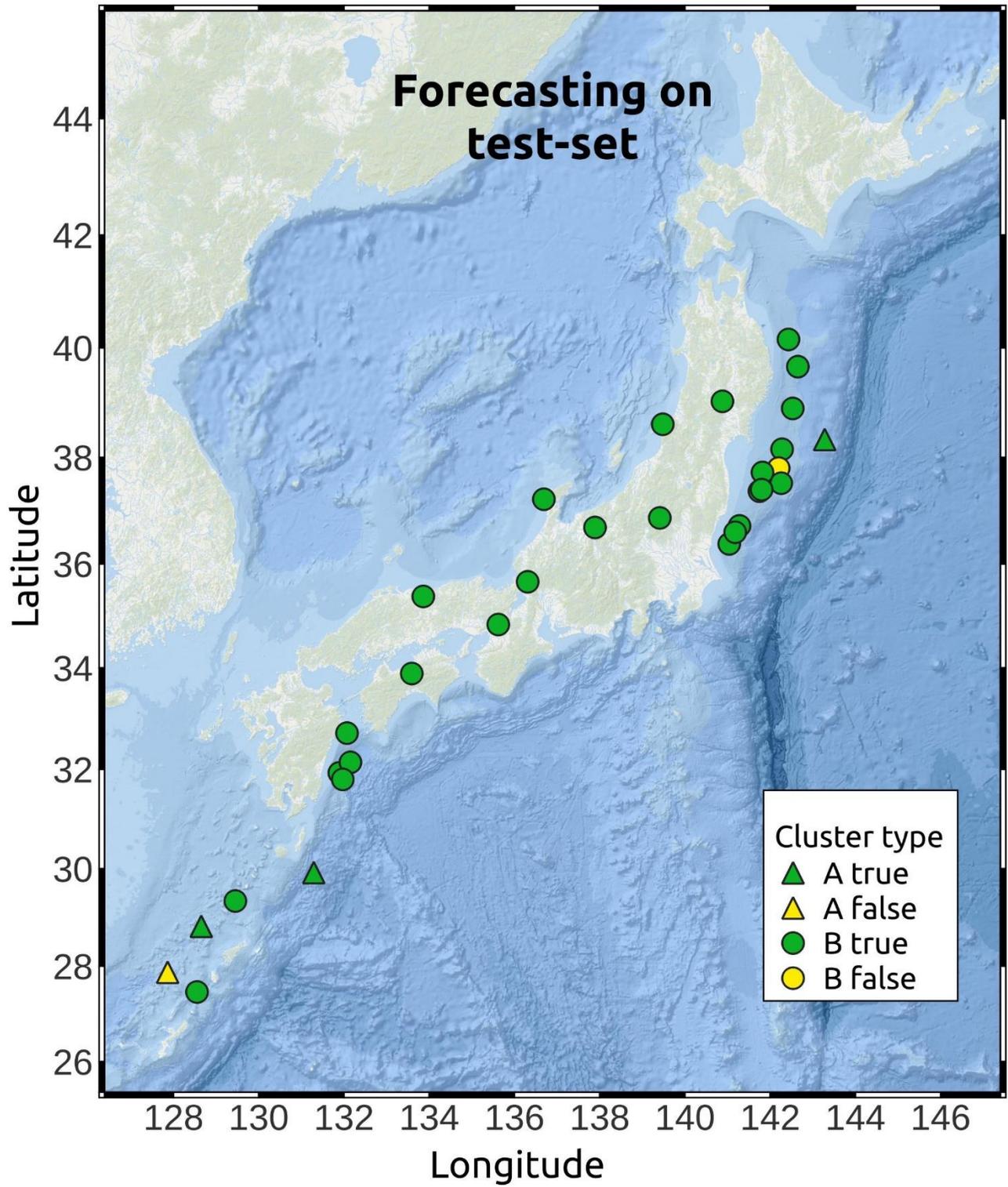

Fig 7a

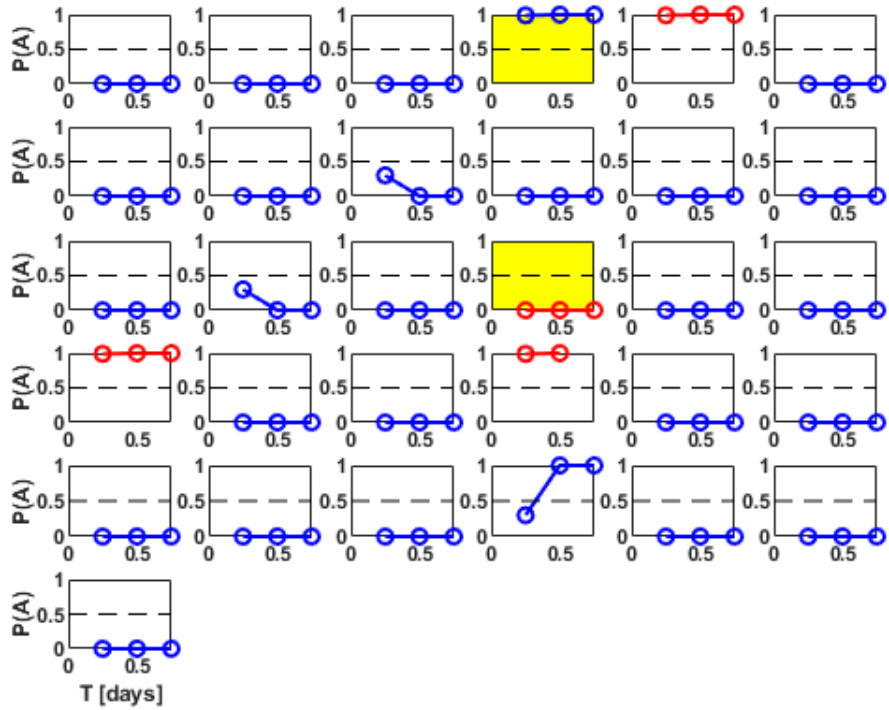

Fig 7b

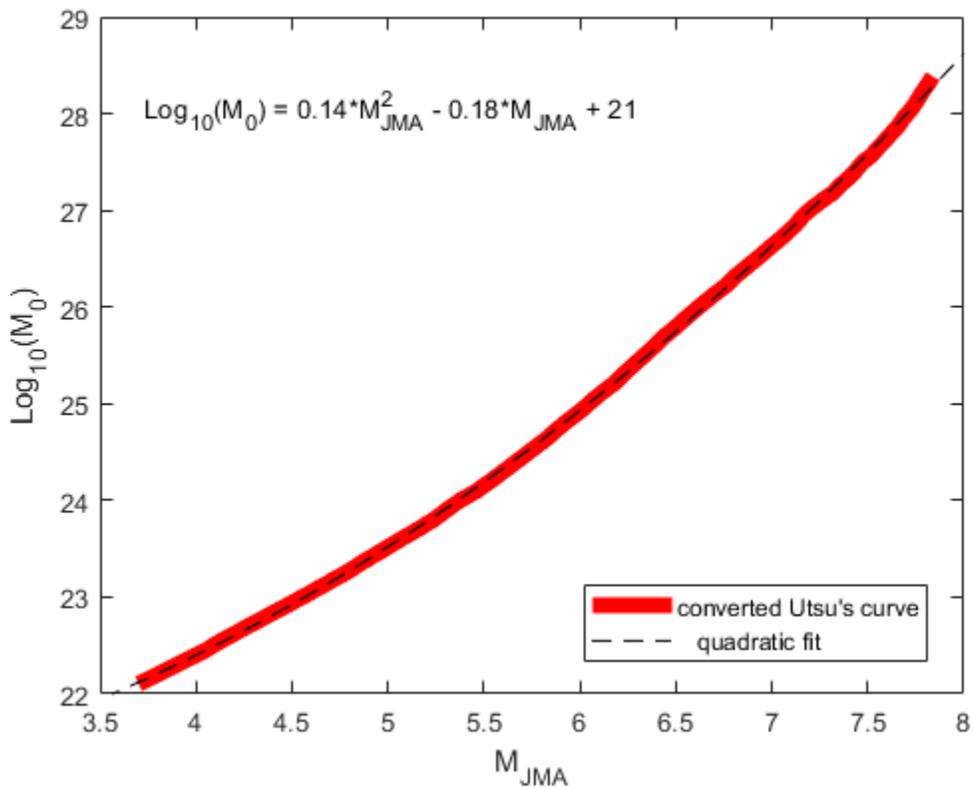

Fig 8

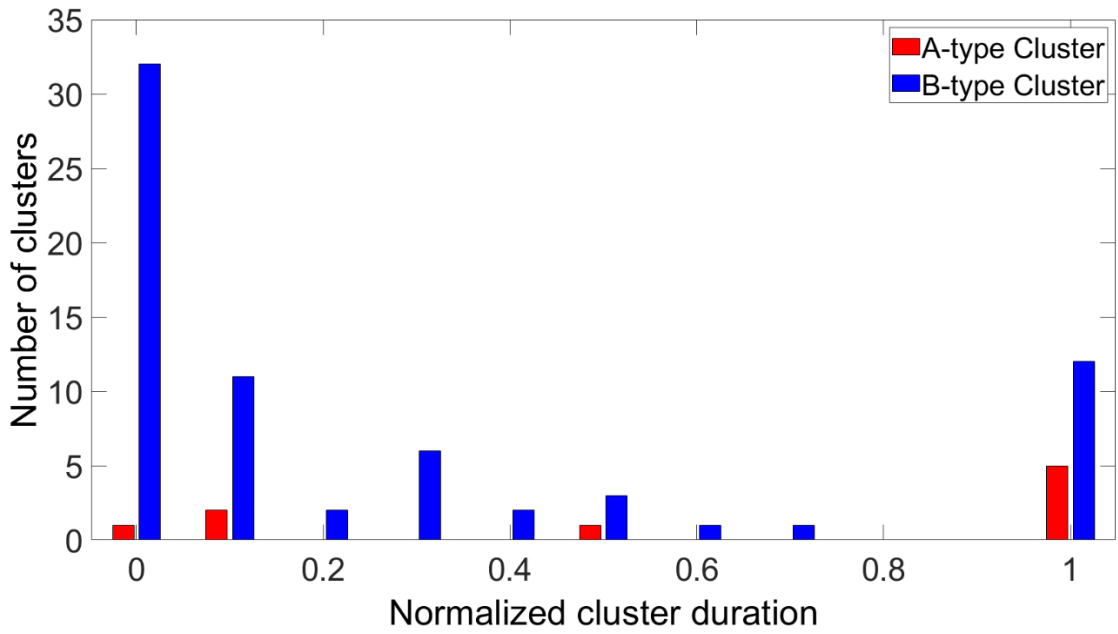

Fig 9a

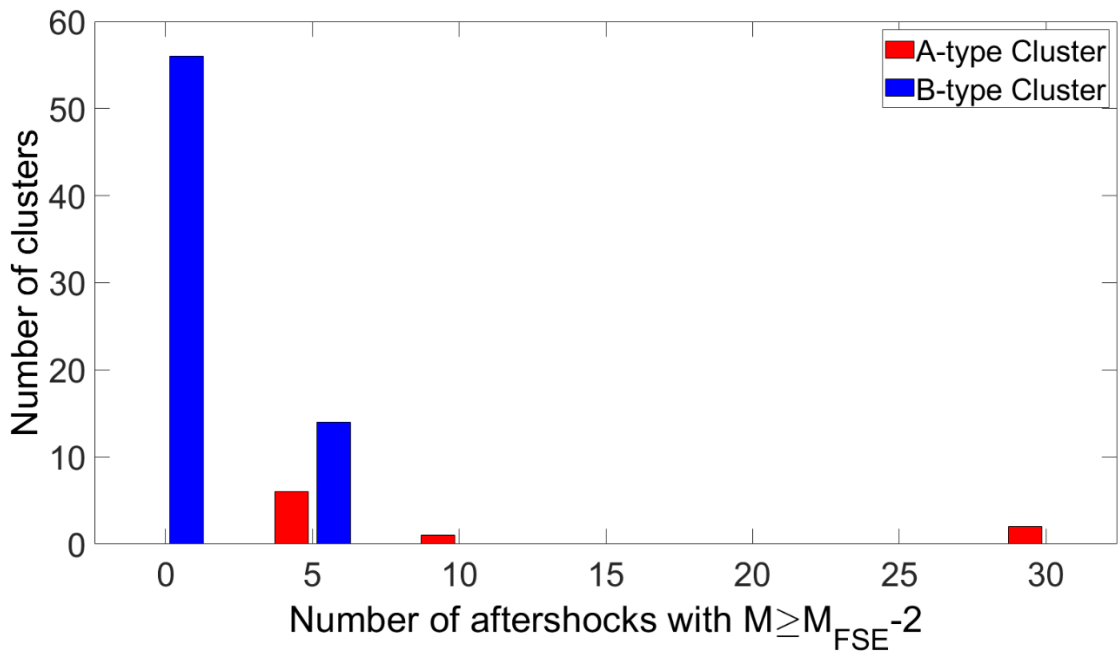

Fig 9b

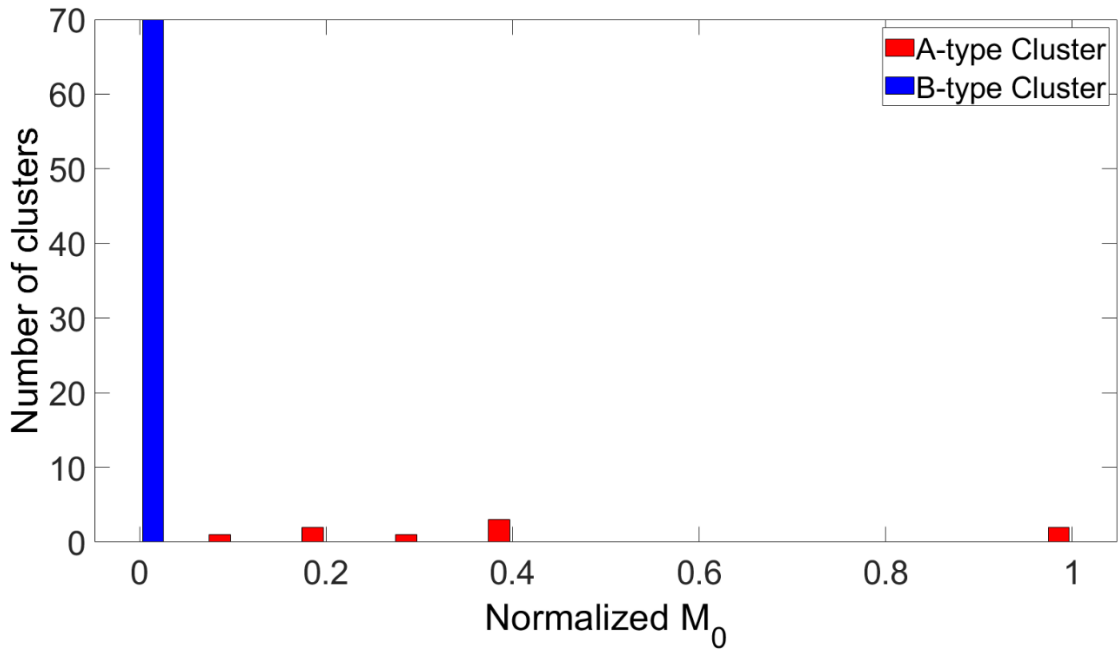

Fig 9c

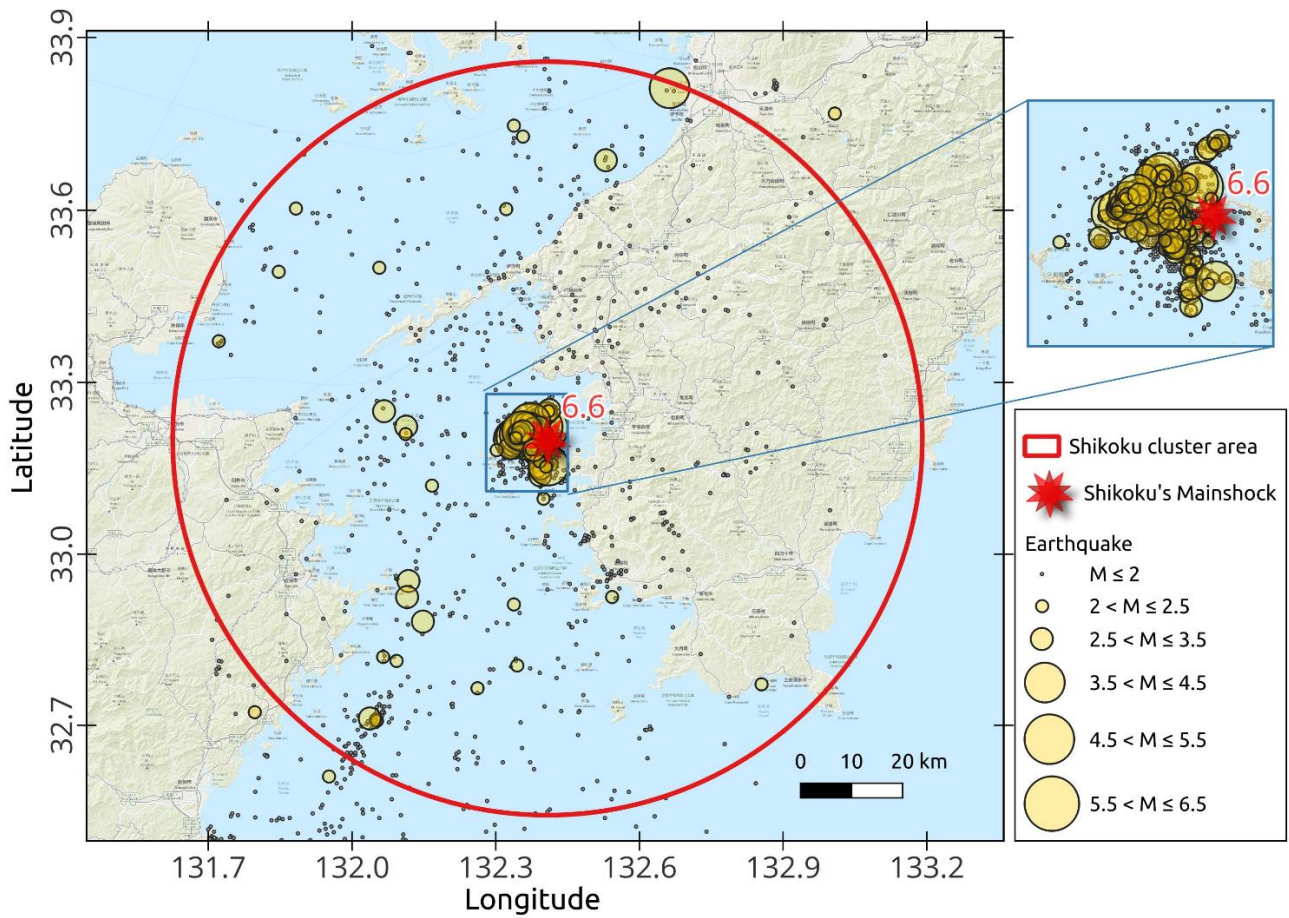

Fig 10a

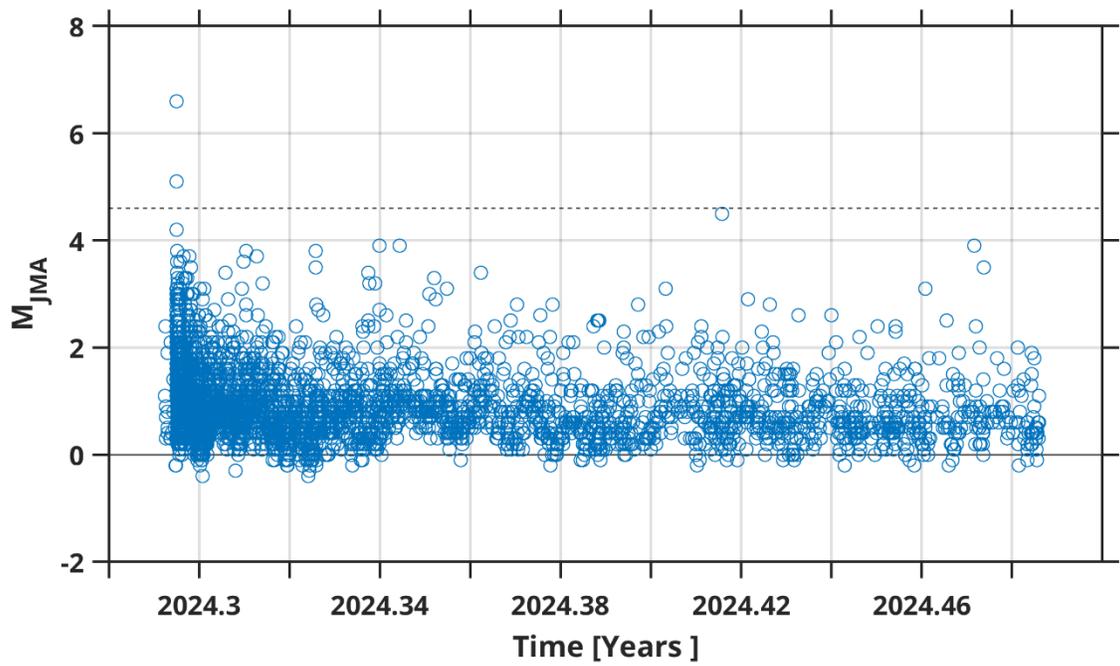

Fig 10b

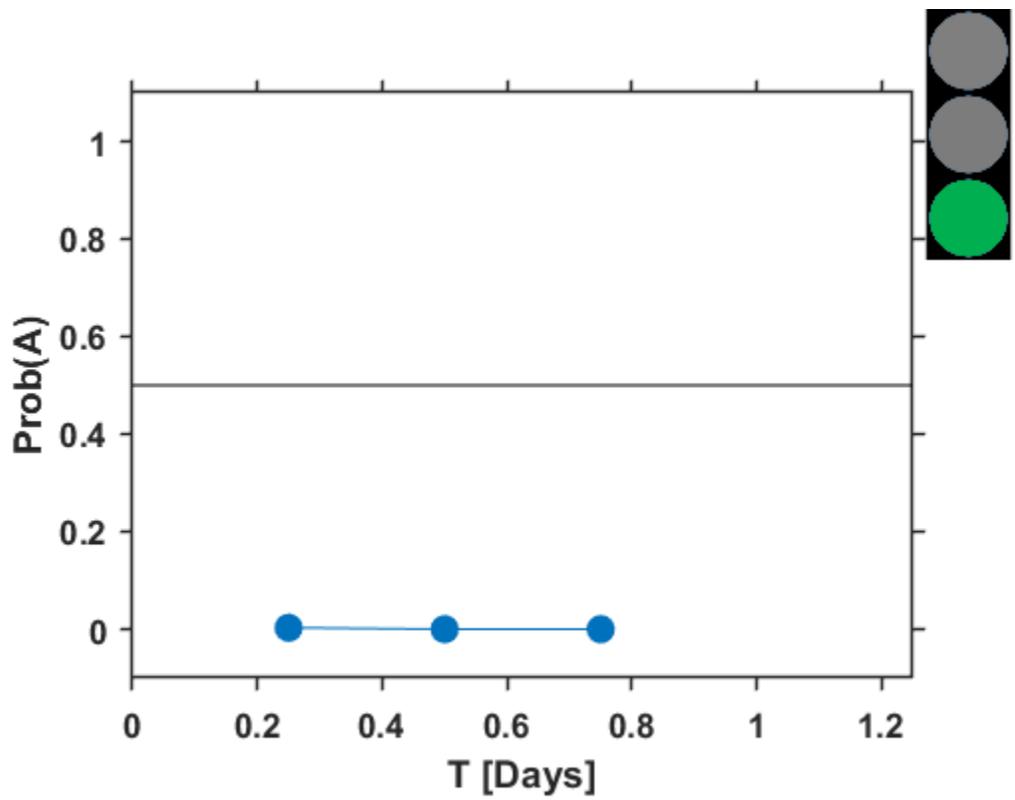

Fig 11

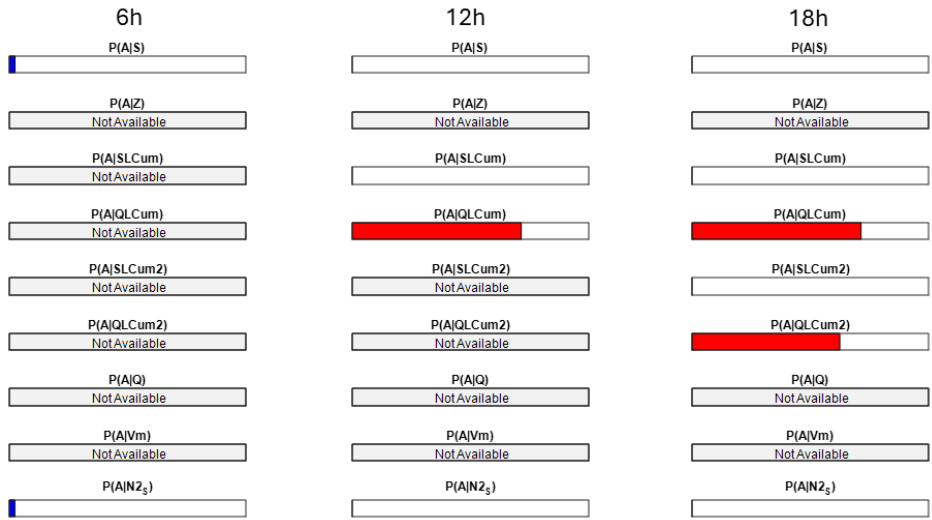

Fig 12a

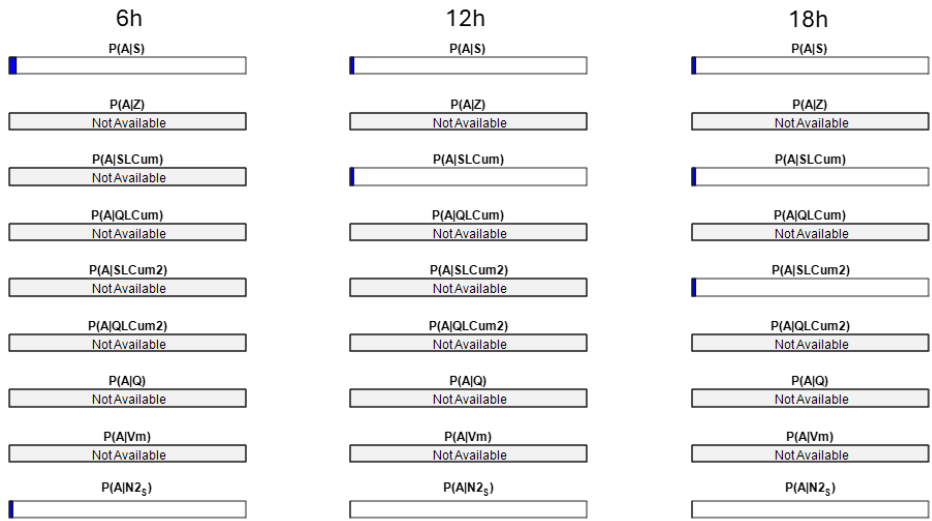

Fig 12b

# 12. Tables

|  | Time intervals (days) | | | | | | | | | |
|---|---|---|---|---|---|---|---|---|---|---|
|  | 0.25 | 0.5 | 0.75 | 1 | 2 | 3 | 4 | 5 | 6 | 7 |
| A-type clusters | 11 | 8 | 8 | 7 | 4 | 4 | 3 | 2 | 2 | 2 |
| B-type clusters | 45 | 45 | 45 | 45 | 45 | 45 | 45 | 45 | 45 | 45 |
| A-type clusters percentage | 20% | 15% | 15% | 13% | 8% | 8% | 6% | 4% | 4% | 4% |

Table 1

|  | Time intervals (days) | | | |
|---|---|---|---|---|
|  | 0.25 | 0.5 | 0.75 | 1 |
| A-type clusters | 7 | 6 | 6 | 5 |
| B-type clusters | 43 | 43 | 43 | 43 |
| A-type clusters percentage | 14% | 12% | 12% | 10% |

Table 2

| Features | Time intervals (days) | | | | Good range (days) | |
|---|---|---|---|---|---|---|
| | 0.25 | 0.5 | 0.75 | 1 | Min | Max |
| S | 0.0415 | 0.0415 | 0.0415 | 0.0415 | 0.25 | 0.25 |
| SLcum | - | 0.0415 | 0.0415 | 0.0415 | 0.5 | 0.5 |
| QLcum | - | 8.046 | 8.046 | 8.046 | 0.5 | 0.5 |
| SLcum2 | - | - | 0.0418 | 0.0418 | 0.75 | 0.75 |
| QLcum2 | - | - | 8.245 | - | 0.75 | 0.75 |
| N2$_S$ | 7.6029 | 7.6029 | 7.6029 | 7.6029 | 0.25 | 0.5 |

Table 3

| Features | Time intervals (days) | | | Good range (days) | |
|---|---|---|---|---|---|
| | 0.25 | 0.5 | 0.75 | Min | Max |
| S | 0.0525 | 0.0525 | 0.0525 | 0.25 | 0.5 |
| SLcum | - | 0.0525 | 0.0525 | 0.5 | 0.5 |
| QLcum | - | - | - | - | - |
| SLcum2 | - | - | 0.0526 | 0.75 | 0.75 |
| QLcum2 | - | - | - | - | - |
| N2$_S$ | 7.6029 | 7.6029 | 7.6029 | 0.25 | 0.5 |

Table 4